\documentstyle[epsf,psfig,aps]{revtex}
\renewcommand{\thefootnote}{\fnsymbol{footnote}}
\begin{document}
\newcommand{\be}{\begin{eqnarray}}
\newcommand{\dlq}{\lq\lq}
\newcommand{\ee}{\end{eqnarray}}
\newcommand{\ben}{\begin{eqnarray*}}
\newcommand{\een}{\end{eqnarray*}}
\newcommand{\beq}{\begin{equation}}
\newcommand{\eeq}{\end{equation}}
\renewcommand{\baselinestretch}{1.0}
\newcommand{\as}{\alpha_s}
\def\eq#1{{Eq.~(\ref{#1})}}
\def\ap#1#2#3{     {\it Ann. Phys. (NY) }{\bf #1} (19#2) #3}
\def\arnps#1#2#3{  {\it Ann. Rev. Nucl. Part. Sci. }{\bf #1} (19#2) #3}
\def\npb#1#2#3{    {\it Nucl. Phys. }{\bf B#1} (19#2) #3}
\def\plb#1#2#3{    {\it Phys. Lett. }{\bf B#1} (19#2) #3}
\def\prd#1#2#3{    {\it Phys. Rev. }{\bf D#1} (19#2) #3}
\def\prep#1#2#3{   {\it Phys. Rep. }{\bf #1} (19#2) #3}
\def\prl#1#2#3{    {\it Phys. Rev. Lett. }{\bf #1} (19#2) #3}
\def\ptp#1#2#3{    {\it Prog. Theor. Phys. }{\bf #1} (19#2) #3}
\def\rmp#1#2#3{    {\it Rev. Mod. Phys. }{\bf #1} (19#2) #3}
\def\zpc#1#2#3{    {\it Z. Phys. }{\bf C#1} (19#2) #3}
\def\mpla#1#2#3{   {\it Mod. Phys. Lett. }{\bf A#1} (19#2) #3}
\def\nc#1#2#3{     {\it Nuovo Cim. }{\bf #1} (19#2) #3}
\def\yf#1#2#3{     {\it Yad. Fiz. }{\bf #1} (19#2) #3}
\def\sjnp#1#2#3{   {\it Sov. J. Nucl. Phys. }{\bf #1} (19#2) #3}
\def\jetp#1#2#3{   {\it Sov. Phys. }{JETP }{\bf #1} (19#2) #3}
\def\jetpl#1#2#3{  {\it JETP Lett. }{\bf #1} (19#2) #3}
\def\epj#1#2#3{    {\it Eur. Phys. J. }{\bf C#1} (19#2) #3}
\def\ijmpa#1#2#3{  {\it Int. J. of Mod. Phys.}{\bf A#1} (19#2) #3}
\def\ppsjnp#1#2#3{ {\it (Sov. J. Nucl. Phys. }{\bf #1} (19#2) #3}
\def\ppjetp#1#2#3{ {\it (Sov. Phys. JETP }{\bf #1} (19#2) #3}
\def\ppjetpl#1#2#3{{\it (JETP Lett. }{\bf #1} (19#2) #3} 
\def\zetf#1#2#3{   {\it Zh. ETF }{\bf #1}(19#2) #3}
\def\cmp#1#2#3{    {\it Comm. Math. Phys. }{\bf #1} (19#2) #3}
\def\cpc#1#2#3{    {\it Comp. Phys. Commun. }{\bf #1} (19#2) #3}
\def\dis#1#2{      {\it Dissertation, }{\sf #1 } 19#2}
\def\dip#1#2#3{    {\it Diplomarbeit, }{\sf #1 #2} 19#3 }   
\def\ib#1#2#3{     {\it ibid. }{\bf #1} (19#2) #3}
\def\jpg#1#2#3{        {\it J. Phys}. {\bf G#1}#2#3}

\begin{flushright}
 DESY 00-149\\
TAUP-2649-2000\\
\today

\end{flushright}

\vspace*{1cm} 
\setcounter{footnote}{1}
\begin{center}
{\Large\bf Has HERA reached a new QCD regime?}
\\[1cm]
{\Large \bf ( Summary of our view )}\\[1cm]

 E. \ Gotsman$^{a}$, E. \ Levin$^{a,b}$, \   M. \ Lublinsky$^{c}$, \ U. \
Maor$^{a}$, \ 
  E. \ Naftali$^{a}$ and  K. \ Tuchin$^{a}$\\  
~~ \\
~~\\
{\it ${}^{a}$ HEP Department, School of Physics and Astronomy } \\ 
{\it Tel Aviv University, Tel Aviv 69978, Israel } \\ 
~~ \\
{\it ${}^b$ Desy Theory, 22603 Hanburg, Germany}\\
 ~~ \\
{\it ${}^{c}$Department of  Physics, Technion,}\\
{\it Haifa, 32000, Israel}\\
~~\\
~~\\
 
\end{center}
\begin{abstract} 

These notes are a summary of our efforts to answer the question in the
title.  Our answer is in the affirmative as: (i) 
HERA data indicate a  large value of the gluon structure function; (ii) no
contradictions with the asymptotic predictions of high density QCD have
been observed; and (iii) the numerical estimates of our model give a
natural description of the size of deviation from the routine DGLAP
explanation. We discuss the alternative approaches and possible new
experiments.
\end{abstract}
\renewcommand{\thefootnote}{\arabic{footnote}}
\setcounter{footnote}{0}

\section{Problems:}

In the region of low $x$ and low $Q^2$  which is now investigated by
HERA we face two challenging problems
:
\begin{enumerate}
\item\,\,\, Matching of ``hard" processes, that can be successfully
described using perturbative QCD (pQCD),  and ``soft" processes, that
should
be described using non-perturbative QCD (npQCD);

\item\,\,\, Theoretical description of high density QCD
(hdQCD). In this kinematic region we expect that the  
typical distances will be small but the parton density will be so large
that a new non perturbative approach needs to  be developed for
dealing with
this system. 
\end{enumerate}

\section{Main idea:}
 The main  physical idea, on which our phenomenological approach is
based  is  \cite{SAT}:

~

{ \large \em
The above two problems are correlated
and the system of partons always passes through the stage of hdQCD 
( at shorter distances ) before it proceeds to the black box limit, which
we
call
non-perturbative QCD and which, in  practice, we describe using old 
fashion Reggeon phenomenology.}

\section{Scales in DIS:}
Since HERA started to investigate  a new kinematic
domain of hdQCD, we want  to find  the most
fundamental phenomena  which are typical of hdQCD as well as npQCD  in
the HERA
experimental data.
 We would  first like  to determine what  the values of
the scales or typical distances at which two transitions

$$ pQCD\,\,\,\,\,\,\, \longrightarrow \,\,\,\,\,\,\, hdQCD\,\,\,\,\,\,\, 
\longrightarrow \,\,\,\,\,\,\, npQCD $$

occur. However, before discussing these transitions let us recall 
that  a photon - hadron interaction at high energy  in the rest
 frame of the target  has two sequential 
stages in time:(i) $\gamma^* \longrightarrow $  hadron system ( $q
\bar q $ - pair ); and then (ii)  this   hadron system ( $q \bar q $ -
pair )
interacts with the target \cite{GRIB}. This time sequence  allows us to
write
the   
cross section for photon-proton interaction in the form:
\begin{equation} \label{EQ1}
\sigma_{tot} ( \gamma^* p ) =  \sum_n
 | \Psi_n|^2 \,\sigma_{tot} ( n, x
)\,,
\end{equation}
where $\Psi_n$ is the wave function of the  hadron ( parton ) system
produced in the first stage of the process.

The first scale we introduce to separate the long and short distances is $     
r^{sep}_{\perp}$. Roughly speaking, the QCD running coupling constant $
\alpha_S( r^2_{\perp} )$ is small  $ \alpha_S( r^2_{\perp})  < 1 $ for $
r_{\perp} < r^{sep}_{\perp}$ , while $ \alpha_S( r^2_{\perp} ) \approx 1$ 
for $r_{\perp} > r^{sep}_{\perp}$.

 The second scale we associate with the transition between the  low parton
density phase  and the high parton density phase of pQCD. This scale
depends
on the energy of the colliding system and can be determined from the
condition
that the packing factor (PF) of partons in  a parton cascade is equal to
unity \cite{SAT}:
\begin{equation} \label{EQ2}
PF \,\equiv\,\kappa\,\,\,=\,\,\,\frac{3\,\pi^2 \alpha_S}{2
Q^2_s(x)}\,\times\,
\frac{xG(x,Q^2_s(x))}{\pi\, R^2}\,\,=\,\,1 ;
\end{equation}
and $r^2_{saturation} = 4/Q^2_s(x)$.

Fig. 1 shows this packing factor in the HERA kinematic region. One can see
 when both $Q^2$ and $x$
are sufficiently small, we really have a dense parton system at HERA.

\begin{figure}
\begin{center}
\epsfxsize=7cm
\leavevmode
\hbox{ \epsffile{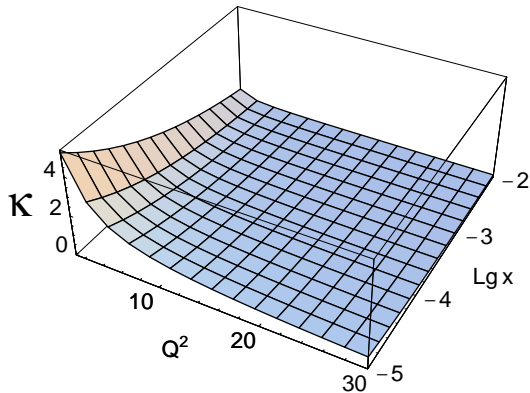}}
\end{center}
\caption{\it The packing factor for the  gluon from HERA experimental
data.}
\label{kappa}
\end{figure}

\section{Theoretical status of high parton density QCD:}

In DIS at low $x$ one can find a high density system of partons,
which is a non-perturbative system due to high density of partons 
although  the
running QCD coupling constant is still small ( $\alpha_S(r_{\perp}) \ll 1
$ ). Such a unique system can be treated theoretically \cite{SAT}. It
should be stressed that the theory of hdQCD is now in  very good shape. 

\begin{figure}
\begin{center}
\epsfxsize=9cm
\leavevmode
\hbox{ \epsffile{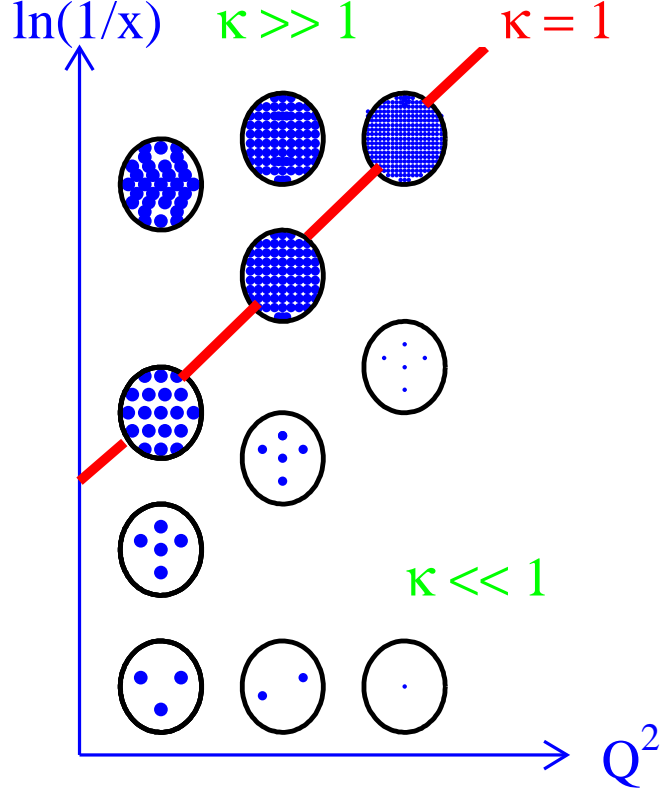}}
\end{center}
\caption{\it The parton distribution in the transverse plane. The curve
shows
the saturation scale $Q_s(x;A)$}
\label{pcsd}
\end{figure}

Two approaches have been developed for hdQCD. The first one
\cite{PTHEORY} is based on pQCD ( see GLR and Mueller and Qiu papers in
Ref. \cite{SAT} ) and on dipole degrees of freedom \cite{MU94}. This
approach gives a natural description of the parton cascade in the
kinematic region for  $\kappa \leq 1$  and up to the
transition region with $\kappa \approx 1 $ ( see Fig. ~\ref{pcsd}).

\begin{figure}
\begin{center}
\epsfxsize=13cm
\leavevmode
\hbox{ \epsffile{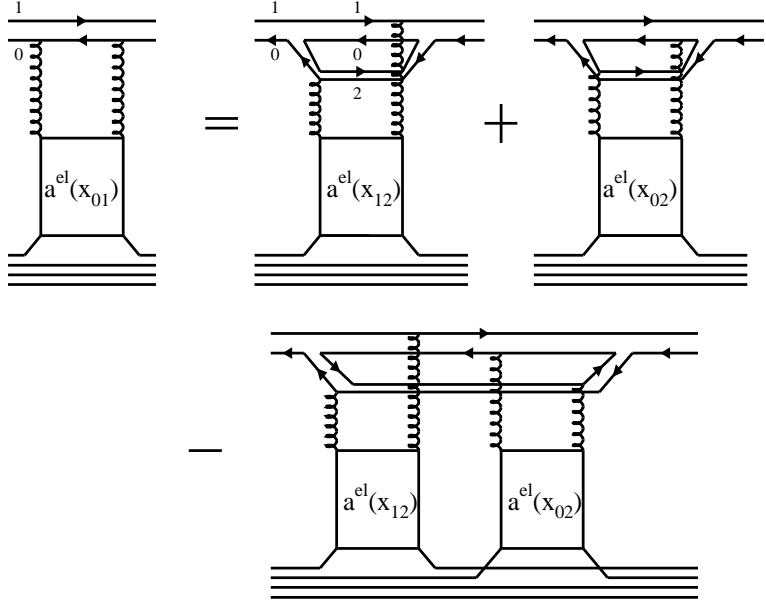}}
\end{center}
\caption{\it The pictorial form of the non-linear evolution equation in
the hdQCD 
kinematic region.}
\label{hdqcd}
\end{figure}

The second method uses the effective Lagrangian suggested by McLerran and
Venugopalan \cite{SAT}, this is a natural framework to describe data 
in the deep
saturation region where $\kappa \gg 1$ (see Fig. ~\ref{pcsd}).
As a result of intensive work using  these two approaches the
non-linear evolution equation has been derived \cite{EQ} which has the
following form
\be \label{GLRINT}   
\frac{d a^{el}({\mathbf{x_{01}}},b_t,y)}{d y}\,\,\,&=&\,\,\,- \,\frac{2
\,C_F\,\as}{\pi} \,\ln\left( \frac{{\mathbf{x^2_{01}}}}{\rho^2}\right)\,\,
a^{el}({\mathbf{x}},b_t,y)\,\,\,+
\,\,\,\frac{C_F\,\as}{\pi}\,\,
\int_{\rho} \,\,d^2 {\mathbf{x_{2}}}\,
\frac{{\mathbf{x^2_{01}}}}{{\mathbf{x^2_{02}}}\,
{\mathbf{x^2_{12}}}}\,\nonumber \\
 &\cdot&\,\,\,
\left(\,\,2\,a^{el}({\mathbf{x_{02}}},{ \mathbf{ b_t -
\frac{1}{2}
x_{12}}},y)\,\,\,-\,\,\,a^{el}({\mathbf{x_{02}}},{ \mathbf{ b_t -
\frac{1}{2}
x_{12}}},y)\,\,a^{el}({\mathbf{x_{12}}},{ \mathbf{ b_t - \frac{1}{2}
x_{02}}},y)\,\,\right)\,\,,
\ee
where $a^{el}(r^2_{\perp},b_t,x)$ is the elastic  scattering amplitude of
the
dipole
of size $r_{\perp}$ at energy $\propto 1/x$ and at impact parameter
$b_t$. The cross section in \eq{EQ1} is equal to $\sigma(r^2_{\perp},x)
=\,2 \,\int \,d^2 b_t \,a^{el}(r^2_{\perp},b_t,x)$. The pictorial form of
\eq{GLRINT} is given in Fig. ~\ref{hdqcd} which shows that this equation
has a very simple physical meaning: the dipole with size $x_{10}$ decays
in two dipoles with sizes $x_{12} $ and $x_{02}$. These two dipoles
interact with the target. The non-linear term which takes into account the
Glauber corrections for such an interaction, \eq{GLRINT} is the same as
the GLR -equation  \cite{SAT} but in the  space
representation. It gives a
correct coefficient in front of the non-linear term which coincides with
one
calculated in Ref. \cite{AGL} in the double log limit. We wish to stress
 that this equation which includes the Glauber rescatterings
, has definite initial conditions and has been derived by both methods
 (see Refs. \cite{EQ,LARRY}). 

We have devoted much time to the  discussion of the pure theoretical
approach as  a comparison of our model approach with the solution to
\eq{GLRINT} \cite{AGL,KOLE} which is the criterion of how well or how
badly our
model works. However, we still need a model because (i) we have to
choose a correct initial distribution to solve \eq{GLRINT}; (ii) to
determine the value of the phenomenological parameters which enter
\eq{GLRINT}  through the initial conditions; and (iii) to find out and
extract the value of the gluon structure function from
$a(r^2_{\perp},b_t,x)$.

\section{Model:}

The first thing we have to specify when constructing a model, is the
choice of the correct degrees of freedom (DOF) or, in other words, we
have to
answer the question  what we mean by letter $n$ in \eq{EQ1}.   

It is well known that for short distances ($r_{\perp}
\,<\,r^{sep}_{\perp}$) the correct degrees of freedom in QCD are
colorless dipoles\cite{MU94}. This means that in \eq{EQ1} at short
distances
\be 
\Psi_n\,\,\,& \longrightarrow &\,\,\,\Psi(Q^2; r^2_{\perp},z)\,\,;
\label{DOF1}\\
\sigma(n,x) \,\,\,&\longrightarrow &\,\,\,\,\sigma(r^2_{\perp},x);
\label{DOF2}
\ee
where $Q^2$ is the photon vituality, $z$ is the fraction of energy
carried by quark and $r_{\perp}$ is the dipole size \footnote{Actually,
\eq{DOF1} and \eq{DOF2} were understood  long before Ref. \cite{MU94}.
In Ref.\cite{DOF1} it was shown that the dipole size is the correct degree
of
freedom for the interaction of quark-anti quark pair at high energy, if
this interaction is induced by two gluon exchange. In Ref. \cite{DOF2}
this
claim was generalized for the leading log approximation of pQCD. In Ref.
\cite{DOF3} it was proved that a dipole picture can be used for describing
the gluon - hadron interaction. However, the final point in Ref.
\cite{MU94} was the most important since it was proven that the QCD
evolution at low $x$ can be rewritten in terms of colorless dipoles.}.

However, at long distances ($r_{\perp} \,>\,r^{sep}_{\perp}$) what the
correct
degrees of freedom are, is still an open question.
 In our approach we use the constituent quarks \cite{AQM}  as the
 degrees of freedom. This assumption leads to
\be 
\Psi_n\,\,\,& \longrightarrow &\,\,\,\Psi_{hadron}( r_i )\,\,;
\label{DOF3}\\
\sigma(n,x) \,\,\,& \longrightarrow &\,\,\,\,\sigma_q(x_i);
\label{DOF4}
\ee 
 where $r_i$  are the quark  coordinates  and $\sigma_q(x_i)$ is the cross
section of the constituent quark $i$ with fraction of energy $x_i$ with
the target.  This assumption is certainly a pure model assumption and an
argument in support of  it is the fact that the same assumption is a part
of the 
successful Regge\cite{DL} and VDM \cite{VDM}  phenomenology  and
that such degrees of freedom
appear in the instanton models of the QCD vacuum \cite{SS}.

Therefore, for long distances the scattering amplitude \eq{EQ1} can be
written
\beq \label{LDF}
\sigma_{tot}( \gamma^* p )  \,\,=\,\,\sum_{M_n \leq
M_0}\,\,\frac{1}{Q^2 + M^2_n}\,\, | \Psi^{\gamma*}_n(r_i)|^2 \,\sum_{q_i}
\sigma_{q_i}(x_i)
\,\,,
\eeq
where $M_0 \approx 1/r^{sep}_{\perp}$. It is clear that \eq{LDF} is
nothing more than the VDM model with AQM prescription for the rescatterings
of vector mesons with the target ( see more details in Ref. \cite{GLMPHP}
).

For short distances ( $r_{\perp} < r^{sep}_{\perp}$ )  \eq{EQ1} reduces
to the form
\beq \label{SDF}
\sigma_{tot}( \gamma^* p )\,\,\,=\,\,\,\int\,\,d^2 r_{\perp} \int \,d
z\,\, 
|\Psi^{\gamma^*}(Q^2; r_{\perp}, z)|^2 \,\,\sigma_{dipole}(r_{\perp},
x)\,\,,
\eeq
where the wave function of the virtual photon are well known
\cite{DOF3,WF}.

For $ \sigma_{dipole}(r_{\perp},x)$ we use the Glauber - Mueller (eikonal
) formula \cite{DOF1,DOF2,DOF3}

\beq \label{GM1}
\sigma_{dipole}(r_{\perp},x) \,\,=\,\,2\,\,\int\,d^2 b_t
\,\left(\,1\,\,-\,\,e^{ - \frac{\Omega(r_{\perp},x;b_t)}{2}}\,\right)\,\,,
\eeq
where
\beq \label{GM2}
\Omega(r_{\perp},x;b_t)\,\,\,=\,\,\,\frac{\pi^2 r^2_{\perp}}{3 \pi R^2}\,x
G(x,  \frac{4}{r^2_{\perp}};b_t)\,\,.
\eeq

It should be stressed that in our calculations we used for $G(x,
\frac{4}{r^2_{\perp}};b_t)$ the Glauber-Mueller formula \cite{DOF3},
namely,
\beq \label{GM3}
xG(x,  \frac{4}{r^2_{\perp}}; b_t)\,\,=\,\,\frac{4 \pi
R^2}{\pi^2}\,\int^1_x
\frac{d x'}{x'} \,\int^{\infty}_{r_{\perp}}\,\frac{ d
r'^2_{\perp}}{r'^4_{\perp}}\,\,2\,\left(\,1 \,\,-\,\,e^{ -
\frac{\Omega^{DGLAP}(r_{\perp},x';b_t)}{2}}\,\right)\,,
\eeq
with
\beq \label{GM4}
\Omega^{DGLAP}(r_{\perp},x;b_t) \,\,=\,\,\frac{3\,\pi^2 r^2_{\perp}}{4 \pi
R^2}\,\,e^{ - \frac{b^2_t}{R^2}}\,\,x G^{DGLAP}(x,
\frac{4}{r^2_{\perp}})\,\,.
\eeq

\begin{table}
\centerline{\bf Table 1}
\begin{tabular}{l l l l}
Reaction & $Q^2\, ( GeV^2) $ & $x$ & References \\
\hline
$\sigma_{tot}(\gamma^* p )$ & $0 \div 65$ & $ < 0.01$ & \cite{GLMPHP} \\
   &  & & \\
$F_2(x,Q^2)$ & $1 \div 65 $ & $< 0.01 $ & \cite{AGL,GLMF2} \\
  &  & & \\
$xG(Q^2,x)$ & $1 \div 65 $ & $< 0.01 $ & \cite{AGL} \\
  &  & & \\
$d F_2/d ln Q^2$ & $ 1 \div 65 $ & $<  0.01 $ & \cite{GLMSLP,OSAKA}\\
  &  & & \\
$\sigma_{tot} ( \gamma\, \gamma^* ) $ & $0; 0 \div 20$ & $ < 0.01$ &
\cite{GLMPHPH}\\
  &  & & \\
$\sigma^{diff}_{tot}$ & $5 \div 65$ & $ < 0.01$ & \cite{GLMINDD} \\
 &  & & \\
$\frac{\sigma^{diff}_{tot}}{\sigma_{tot}}$ & $1 \div 65$ & $ < 0.01$
&\cite{GLMRDT}\\
 &  & & \\
$\sigma( \gamma^* p \rightarrow J/\Psi + p ) $ & $0  \div 65$ & $ < 0.01$
& \cite{GLMVP,OSAKA} \\
 &  & & \\
slope $B (  \gamma^* p \rightarrow J/\Psi + p ) $ & $0  \div 65$ & $ <
0.01$   &  \cite{GLMVP,OSAKA}\\
 &  & & \\
slope $B (  \gamma^* p \rightarrow \rho  + p ) $ & $5  \div 65$ & $ <    
0.01$   &  \cite{GLMVP}\\
\end{tabular}
\end{table}

In \eq{GM4} we use the Gaussian parameterization for the profile function
for the quarks in the proton
\beq \label{GM5}
S(b_t) \,\,=\,\,\frac{1}{\pi R^2} \,e^{ - \frac{b^2_t}{R^2}}\,\,,
\eeq
where $R$ is the proton radius.

\eq{GM1} - \eq{GM5} take into account the rescattering quark - anti quark
pair and one ( the fastest) gluon in the parton cascade. It turns out that
the rescattering of the fastest gluon is very important in our attempts to
describe the experimental data. The data, which we described, as well as
the details of our model,  can be found in references, presented in Table
1.

The main features of our model are  summarized in the following Table2.

~

~
\begin{table}
\centerline{\bf Table 2}
\begin{tabular}{ l l l}
 {\large Perturbative QCD} & $\longrightarrow$ &
{\large  non-perturbative QCD}\\
 & & \\
{\large short distances} &  $\longrightarrow$ &  {\large long distances}
\\
 & & \\
{ \Large $ r_{\perp} \,\,<\,\,$} & {
 \Large $ r^{sep}_{\perp}$}
&\,\,\,\,\,\,\,\,\,\,\,\,\,\,\,\,{  \Large
$<\,\,
r_{\perp}  
$}\\
   & & \\
DOF: color dipoles \cite{MU94}& $\bullet$ & DOF: constituent
quarks\cite{AQM} \\
 &  & \\
 $\Psi_n$:  QED for virtual photon  & $\bullet$ & $\Psi_n$: generalized
VDM for
hadronic system\\ 
 & &\\
$\sigma_{tot}(n,x) = \sigma (r^2_t,x)$ & $\bullet$ &
$\sigma_{tot}(n,x) =
\sigma(q q \rightarrow q q; x)$ \\
 & & \\
Mueller-Glauber Eikonal \cite{DOF3}  for $\sigma (r^2_t,x)$ & $\bullet$ &
Regge phenomenology
for $\sigma(q + q \rightarrow q + q; x)$\\
\end{tabular}
\end{table}

\begin{figure}[hptb]
\begin{tabular}{ c c}
\psfig{file=diptot.ps,width=70mm} & \psfig{file=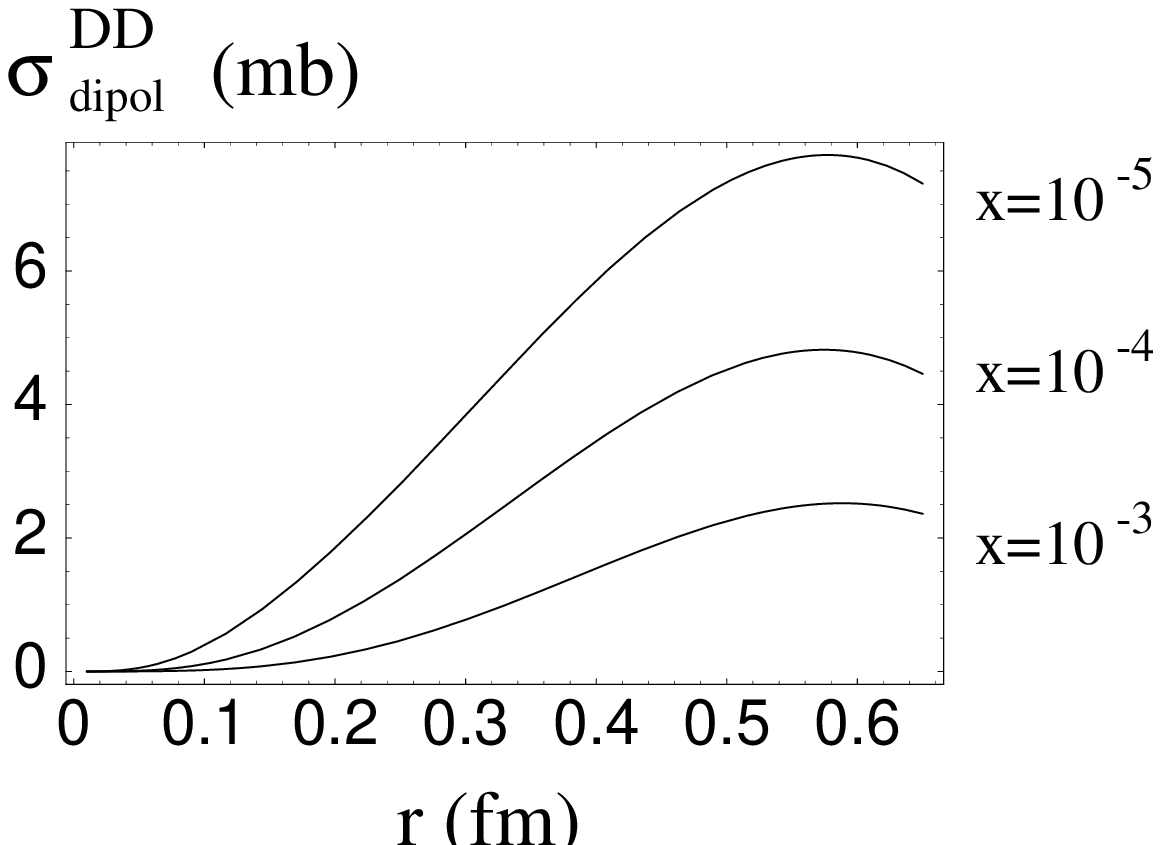,width=70mm}\\
       &  \\
Fig.4-a & Fig. 4-b \\
 &  \\
\end{tabular}
\caption{\it  The $r_{\perp}$ dependence of $\sigma^{dipole}_{tot}$
(Fig. 4-a) 
and
$\sigma^{dipole}_{diff}$ (Fig.4-b).}
\label{dipsi} 
\end{figure}

Fig.~\ref{dipsi}  shows the $r_{\perp}$ dependence of the total and
diffractive dipole cross sections, where $\sigma^{dipole}_{tot}$ is
calculated using \eq{GM1} - \eq{GM5} and  where the diffractive cross
section for a dipole is estimated by the Kovchegov-McLerran formula
\cite{KM} 
$$
\sigma^{dipole}_{diff}\,\,=\,\,\int\,d^2\,b_t\,\,\left(\,1 \,\,-\,\,e^{-
\frac{\Omega}{2}}\,\right)^2 $$
with $\Omega$ defined by \eq{GM2} - \eq{GM5}.

\section{ Advantages  of the model:}
 \begin{itemize}

\item\,\,\,For short distances $r_{\perp} < r_{saturation} \approx
1/Q_s(x)$ the model agrees with the pQCD predictions. They are the same as
in
DGLAP evolution equations;
\item\,\,\, The model gives a natural explanation and an estimate for the
value of the saturation scale (see Fig.~\ref{sat});
  
\begin{figure}
\begin{center}
\epsfxsize=7cm
\leavevmode
\hbox{ \epsffile{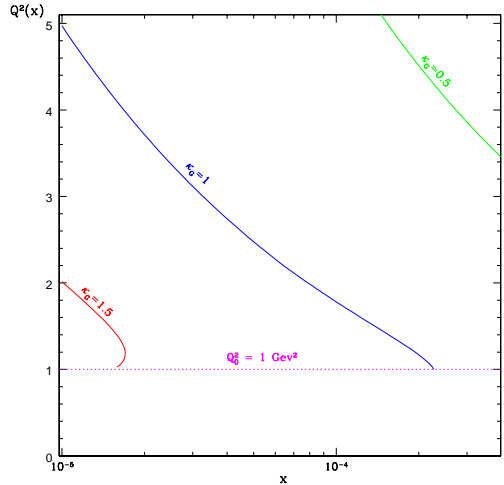}}
\end{center}
\caption{\it The saturation scale $ Q_s(x)$ in our model .}
\label{sat}
\end{figure}

\item\,\,\,This model is a good approximation to the full solution of the
 nonlinear evolution equations that describe theoretically the
hdQCD kinematic region ( GLR -equation and its' generalizations
\cite{SAT,PTHEORY,ELTHEORY,EQ}). See Figs.6-a - 6-c  and discussion below;

\item\,\,\, Our model correctly reproduces the operator product  expansion
 (OPE)  and 
gives the higher twist contributions with the  correct anomalous
dimension\cite{ADHT} in
the region of low $x$;

\item\,\,\,In the model  as well as in Mueller-Glauber approach  
  in general one preserves
 the relation between elastic and
quasi-elastic scattering and multi particle production in DIS based on the
AGK cutting rules\cite{AGK}. This  allows us to describe the inclusive
cross
sections
as well as the  correlation functions;

\item\,\,\, Our model includes the impact parameter behavior of the
dipole-target amplitude, consequently it can be generalized to 
DIS with nuclei. We have completed some calculations for DIS with a 
nuclear
target\cite{GLMA}  but will not discuss them here;

\item\,\,\, The model violates the energy sum rules but 
the discrepancy is very small
as  can be seen in Fig. 6-d;

\begin{figure}[hptb]
\begin{center}
\begin{tabular}{ c c}
\psfig{file=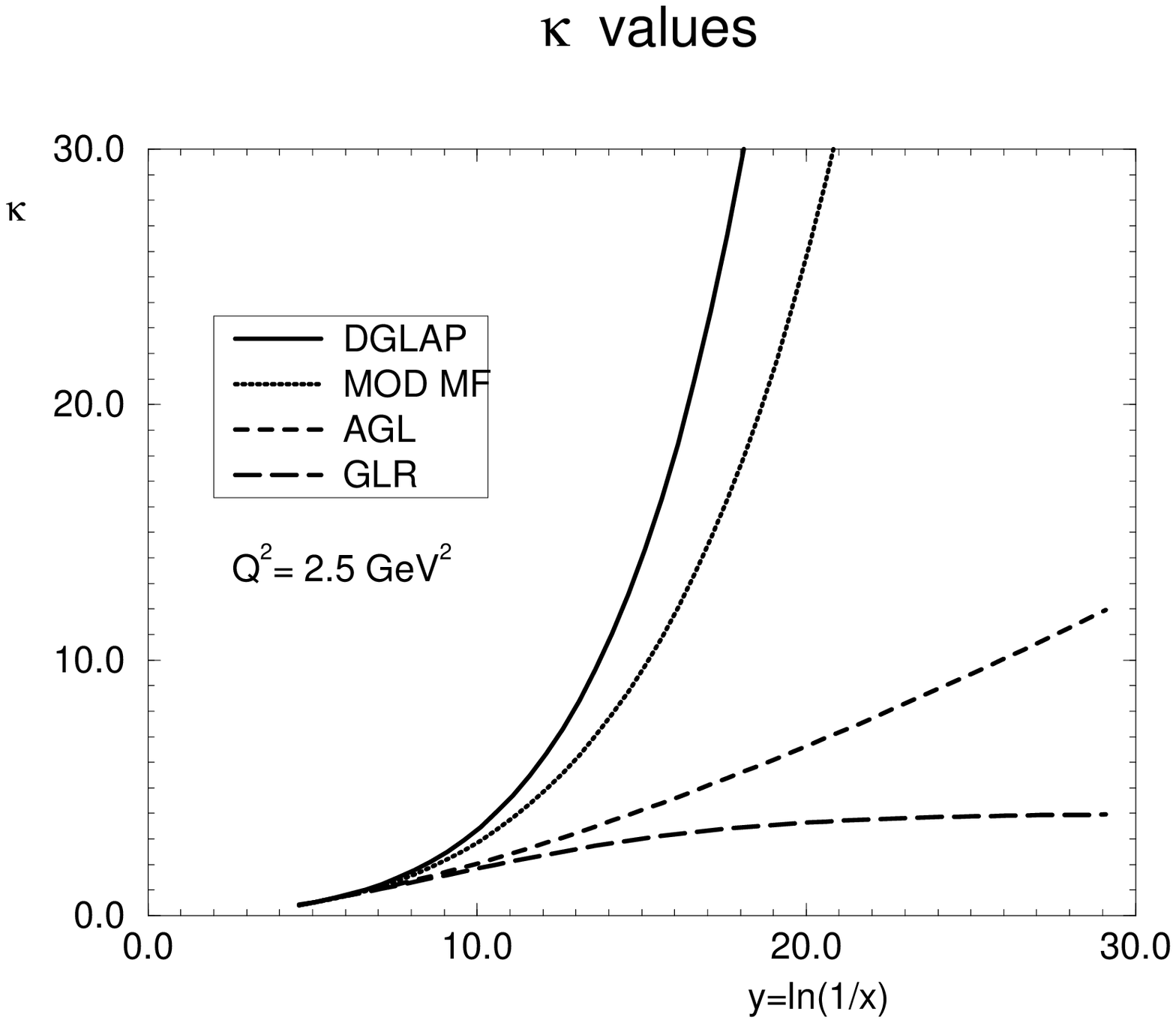,width=70mm} & \psfig{file=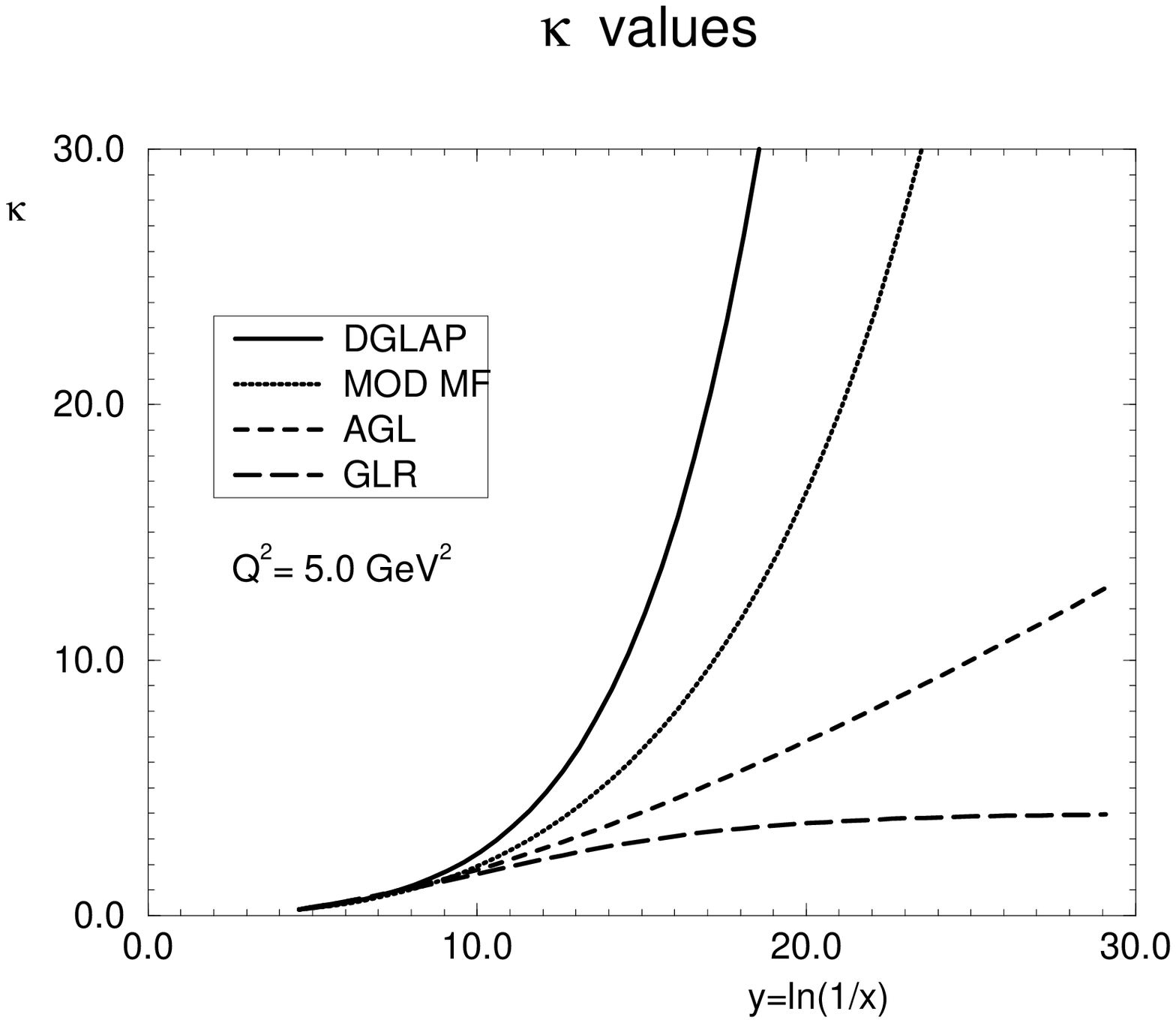,width=70mm}\\
Fig.6-a & Fig. 6-b \\
& \\
 & \\
\psfig{file=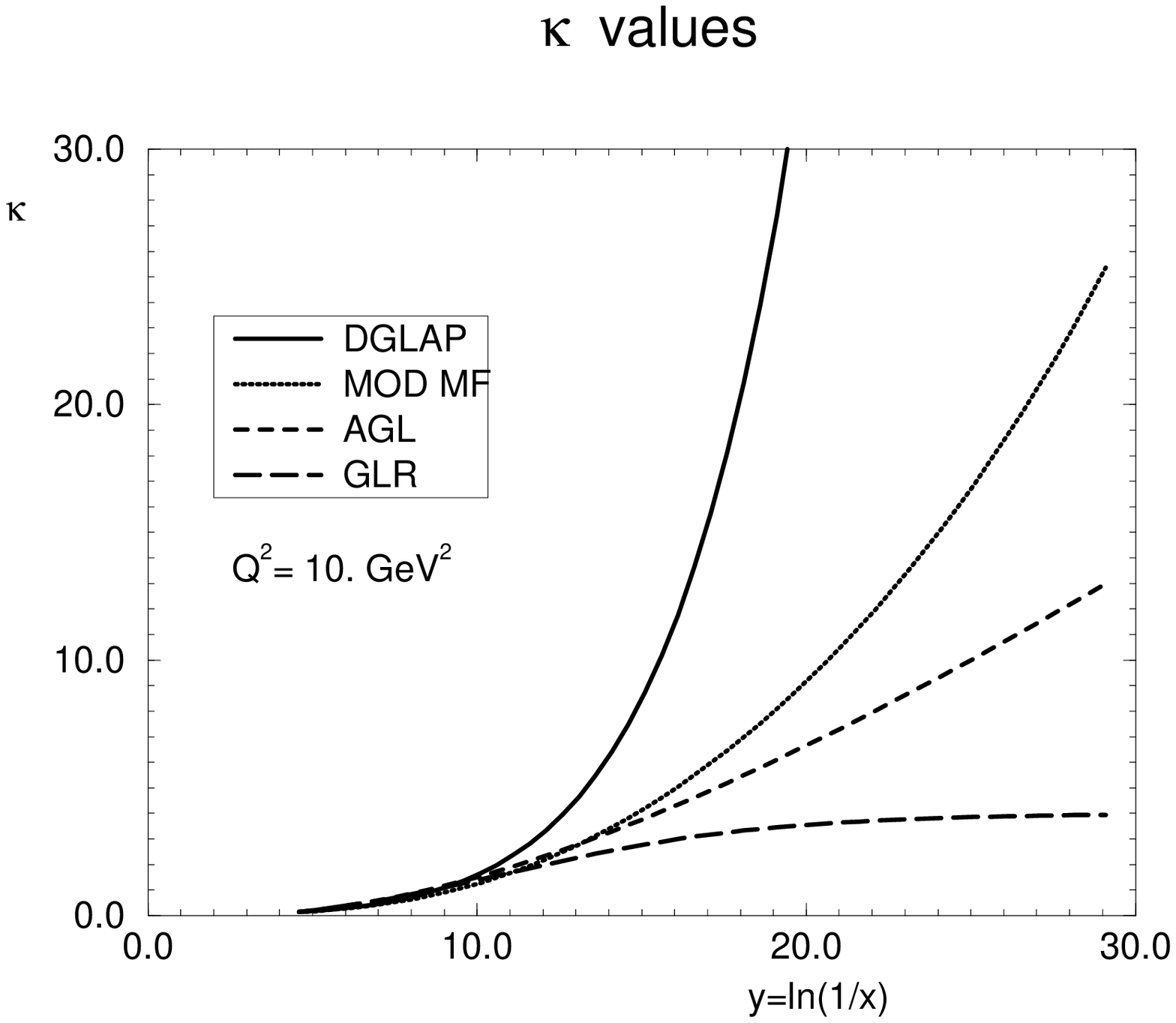,width=70mm} &\psfig{file=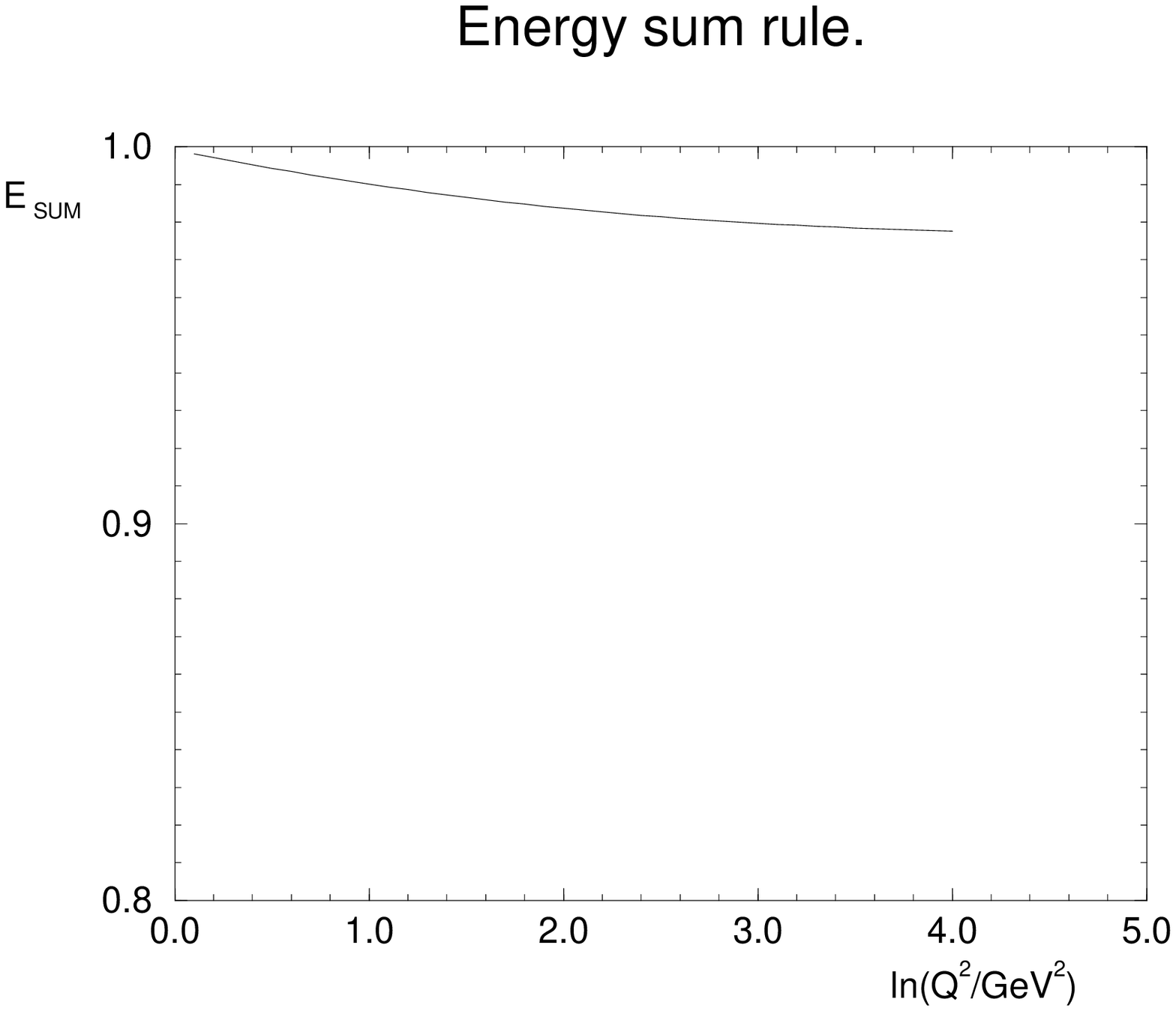,width=70mm}
 \\
Fig. 6-c & Fig. 6-d \\
\end{tabular}
\end{center}
\caption{\it Figs.6-a - 6-c show the packing factor $\kappa$ for gluons
in
the DGLAP evolution
equation (DGLAP), in Mueller-Glauber Eikonal approach (MODMF), in the
correct non-linear evolution equation  (AGL) and in the simplified
version of the GLR equation (GLR). Fig.6-d displays the energy-momentum
sum
rules in our model.}
\label{nlkap}
\end{figure}

\item\,\,\, In the region of large $r_{\perp} > r^{sep}_{\perp}$
the model displays all the  attractive features of the VDM and is  able to
describe ``soft" physics in DIS well. 

\end{itemize}

\section{ Disadvantages of the model:}

We have two major problems with this kind of a  model. First, the
theoretical accuracy of the general formula

\begin{equation} \label{EQ3}
\sigma_{tot} ( \gamma^* p ) =  \int \,\,d^2 r_{\perp}\,dz
\,\,|\Psi^{\gamma^*}( Q^2; r_{\perp},z)|^2 \,\,\sigma_{dipole}(
r^2_{\perp}, x )
\end{equation}
is only approximate. This formula can be proven\cite{DOF2,DOF3,MU94}  only
in the
 leading log(1/x)
approximation of pQCD in which we consider $\alpha_S\,\ln(1/x) \approx\,1$
while $\alpha_S \,\ll\,1$. Therefore, \eq{SDF} is much less general than
the 
DGLAP approach and our results are always worse than those for the DGLAP 
evolution equations.   This is the   price we pay for a  far more
transparent
formalism which
 includes the long distance physics. 

Second, the separation scale $r^{sep}_{\perp}$ changes for
different reactions. Even for longitudinal and transverse polarized photon
we have  different $r^{sep}_T \approx 1/( 0.85 \div 0.95) GeV$
while $r^{sep}_L \geq 0.6 GeV$. This fact  limits the  use of
the attractive factorization properties of \eq{EQ1}  or /and \eq{SDF}.

Other shortcomings:
\begin{itemize}
\item\,\,\, In the region of low $Q^2$  the model's results 
are always less than those found by solving
 the non-linear evolution equation (see Fig.6-a). Therefore, the 
 estimates obtained from the model are below  the measured effects. 
 Knowing these limitations we plan to
improve our model;
\item\,\,\, The eikonal approach cannot be correct at very high energies
so
our model is specially constructed and suited for the HERA kinematic
region
and can only be used with a lot of caution to higher energies and, in
particular, to the LHC energy range;

\item\,\,\, We used a Gaussian profile function in impact parameter
($b_t$)
which  cannot describe the region of very large momentum transfer.
Therefore,  to be accurate we have  to try alternative functions for
large
$t$ processes.  

\end{itemize}

\section{Asymptotic predictions:}
In our model the dipole cross section at high energy and fixed $r_{\perp}$ 
is equal to
\begin{equation} \label{EQ4}
\sigma^{dipole}(r^2_{\perp},x) \,\,=\,\,2 \,\pi < b^2_t(x)>
\end{equation}
where $<b^2_t>$ is the average impact parameter in dipole-target
scattering . As function of energy it behaves in our model   as 
$$< b^2_t > \,\,=\,\,R^2 ( C\,\,+\,\,\ln [Q^2_s(x)r^2_{\perp}]
)\,\,\,\longrightarrow{}_{x\,\rightarrow \,\,0}\,\,\,
2\,\alpha'_{eff}\ln(1/x)\,\,\,\,,$$
 where
$R^2$ is the target size and $C$ is the Euler constant.
Fig. 7-a and Fig.7-b show that the effective slope is rather small ( it is  
at least four
 times smaller than  the soft slope of Pomeron trajectory).
However, for the processes induced by the gluon structure function this
effective shrinkage could be visible and it has to be taken  into account.

\begin{figure}[hptb]
\begin{center}
\begin{tabular}{ c c}
\psfig{file=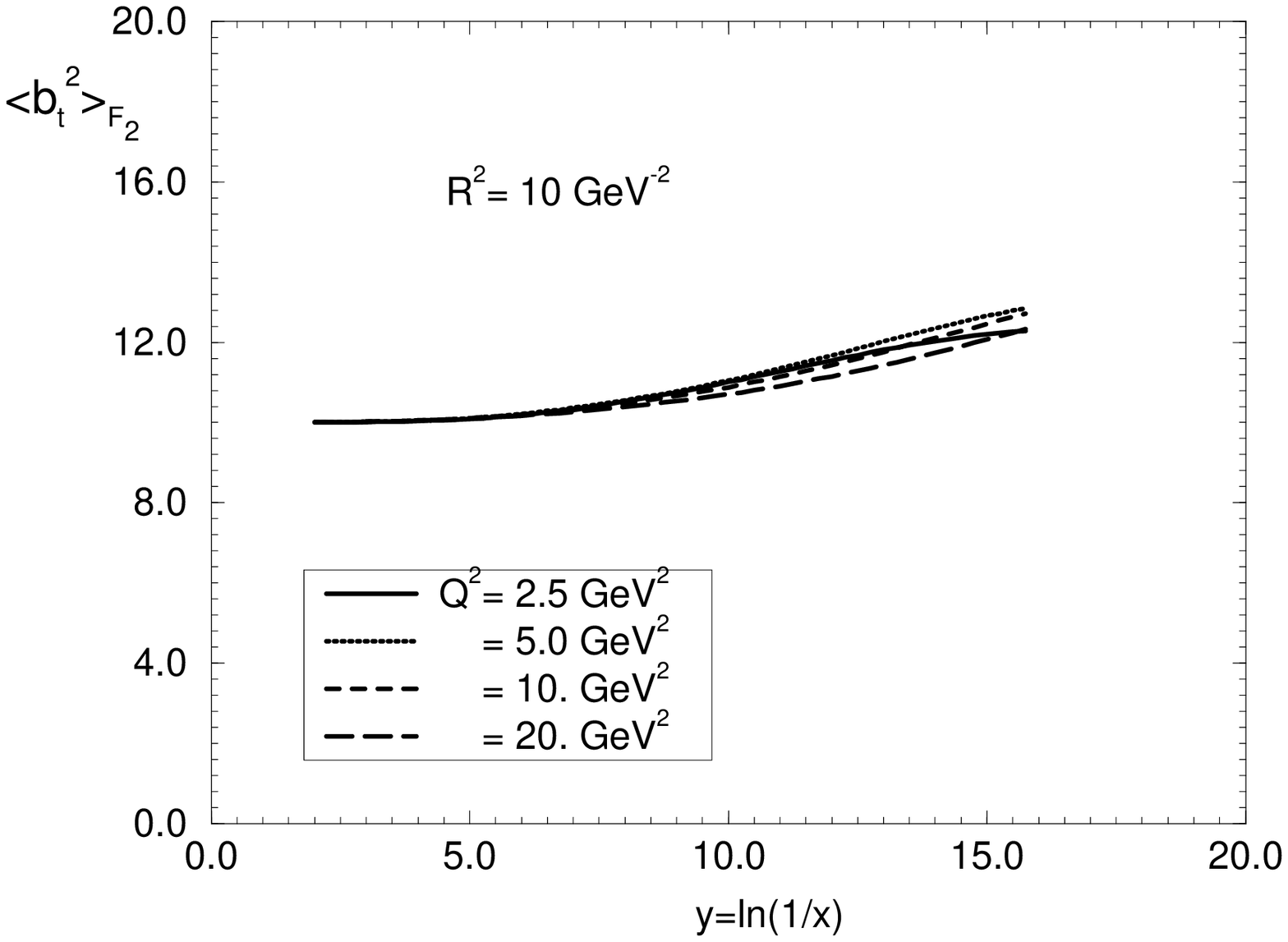,width=70mm} & \psfig{file=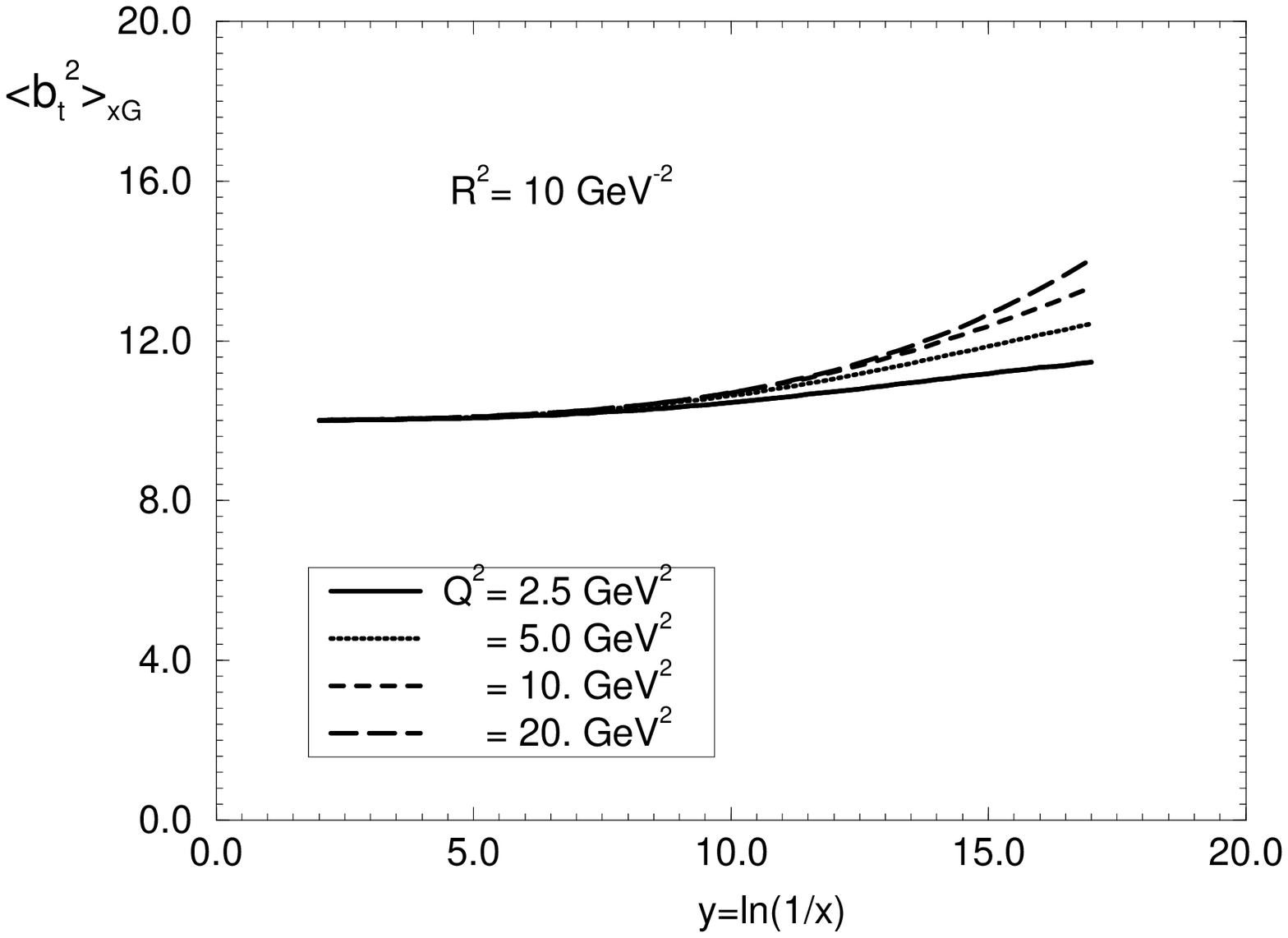,width=70mm}\\
Fig.7-a & Fig. 7-b \\
\end{tabular}
\end{center}
\caption{\it Figs.7-a - 7-b show the effective  shrinkage of the
diffraction peak in our model for $F_2$ and $xG$ 
}
\label{b}
\end{figure}

Such a behavior of the dipole cross section results in the following
predictions for the observables:
\begin{itemize}

\item\,\,\,$F_2\,\longrightarrow\,\,\frac{N_c}{6 \pi^2}
\sum^{N_f}_{1}
Z^2_f\, Q^2\,\cdot\,< b^2_t >
\,\longrightarrow\,\,\frac{N_c}{6 \pi^2} \sum^{N_f}_{1} Z^2_f\,Q^2\cdot
\left( R^2 \,\,+\,\,2\,\alpha'_{eff}
\ln(1/x)\,\right);$
\item\,\,\,\,$\frac{dF_2}{d \ln
Q^2}\,\longrightarrow\,\frac{N_c}{6 \pi^2} \sum^{N_f}_{1}
Z^2_f\,\,Q^2\,\cdot\, R^2( \ln (Q^2_s(x)/Q^2))\,\longrightarrow
\,\,F_2 \frac{ \ln (Q^2_s(x)/Q^2)}{1 + \ln(Q^2_s(x)/Q^2)};$
\item \,\,\, The ratio of the diffractive  to the total cross section  
should
not depend on energy\cite{KM}:
$$\frac{\sigma^{diff}}{\sigma_{tot}}\,\,=\,\,Const\,\,
\stackrel{ slowly}{\longrightarrow} 
\frac{1}{2}\,;$$
\item\,\,\,The energy behaviour of the diffractive cross section is
determined by short distances $r_{\perp}\, 
\approx  \,r^{saturation}_{\perp} = 1/Q_s(x)$;
\item\,\,\,The high density QCD effect should be stronger in
the diffractive
channels. Therefore, we expect to see these effects firstly in diffractive
production of heavy mesons or in inclusive diffractive production;
\item\,\,\, We expect minima in $t$ distribution of the diffractive
production (especially in the case of diffractive production of mesons in
DIS ) which is related to wave propagating picture described by eikonal
approach. 

 \end{itemize}

\section{Phenomenological parameters of the model:}

Before discussing the application to HERA data we  list the
parameters that we use to fit the data.

~

\subsection{ $\mathbf{ R^2}$ - size of the target.}

The size of the target enters the impact parameter profile of the target
which we take in the Gaussian form:
\begin{equation} \label{M1}
S(b_t) \,\,=\,\,\frac{1}{\pi R^2}\,e^{ - \frac{b^2_t}{R^2}}\,\,.
\end{equation}
The HERA data for photo production of J/$\Psi$ - meson as well as CDF data
on double parton cross section leads to the value of $R^2 = 5 \div
10\,\,GeV^{-2}$.  We use $R^2 $ = 5 $GeV^{-2}$ and $R^2 $ = 10 $GeV^{-2}$  
 for an estimate of the possible effect and $R^2$ as a fitting parameter
for the description of the experimental data. Note, that the value of
$R^2= 8.5\,GeV^{-2}$ was taken  for all reactions that we have
described. 

\subsection{ $\mathbf{ Q^2_0 = 1/r^2_{sep}}$ - separation
parameter.}

As we have discussed we can trust our model for  the saturation effect (
see
\eq{GM1} - \eq{GM5} )   only at rather small distances
($r_{\perp} \,<\,r^{sep}_{\perp}$ ) or , in other words, at large
virtualities of the incoming photon $Q^2 > Q^2_0$. We have commented on
the
value of  $r^{sep}_{\perp}$, but in practice we used $Q^2_0 = 0.6
\div  1
\,\,GeV^2$ and tried to study how our fit depends on the value of $Q^2_0$.
Therefore, the result of our calculations should be read correctly, as
{\it ``the shadowing  corrections from short  distances $r_{\perp} <
1/Q^2_0$ gives this or that ...."}.

\subsection{ Solution of the DGLAP  evolution equations.}

We tried to use all available parameterization of the solution of the
DGLAP evolution equations\cite{CTEQ,MRS}, but we  prefer  the GRV
parameterization\cite{GRV} . The reason for this is very simple: the
theoretical
formulae, that are  the basis of our model,  were derived in double log
approximation of pQCD and the GRV parameterization is  the closest one to
the DLA.

\section{Our model versus HERA data:}
\subsection{$\mathbf{\sigma_{tot}(\gamma^* + p )}$ at low $\mathbf{x} $
and $\mathbf{Q^2}$.}
\begin{figure}
\begin{tabular}{l l}
\psfig{file=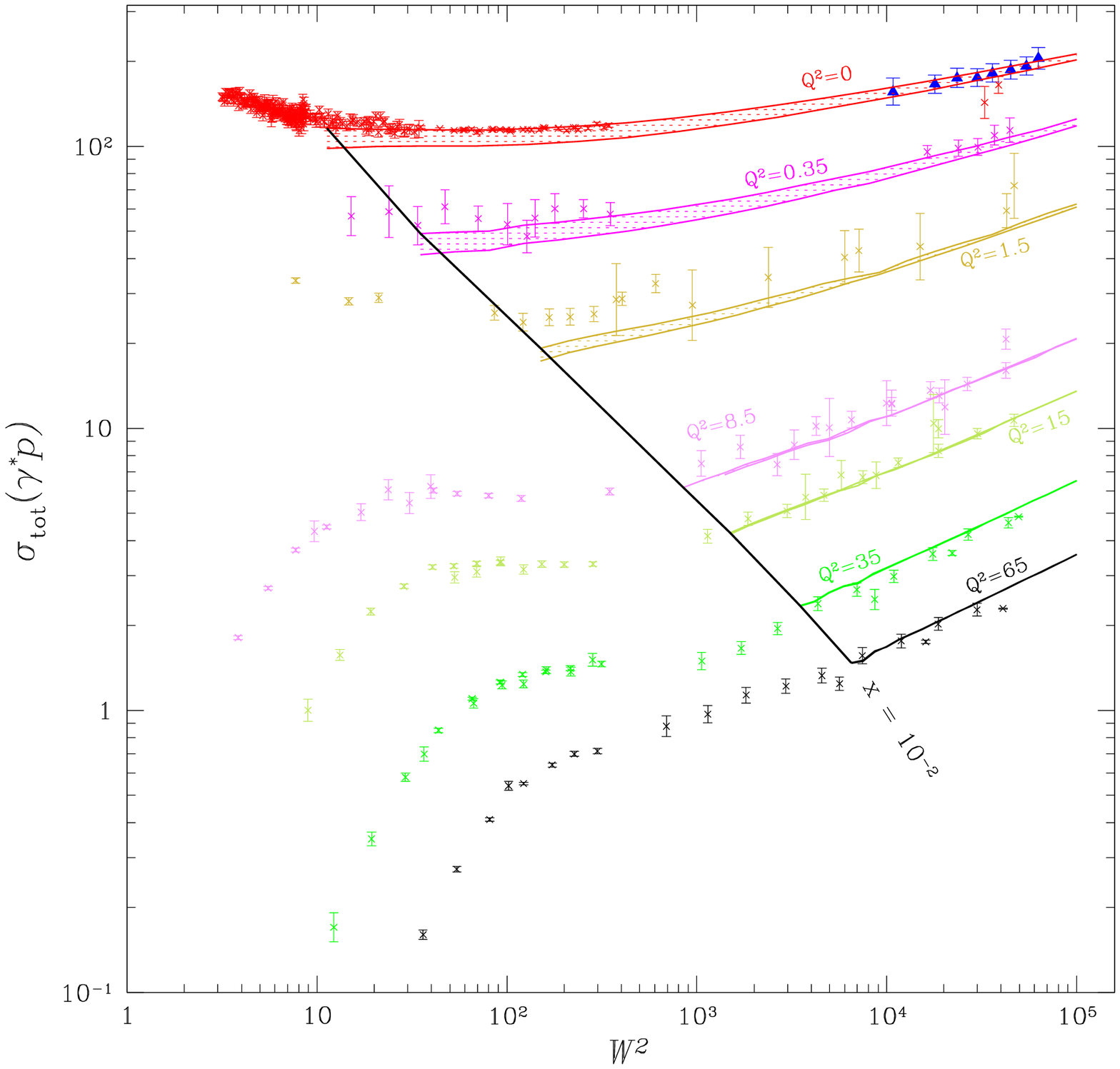,width=8.5cm} &
\psfig{file=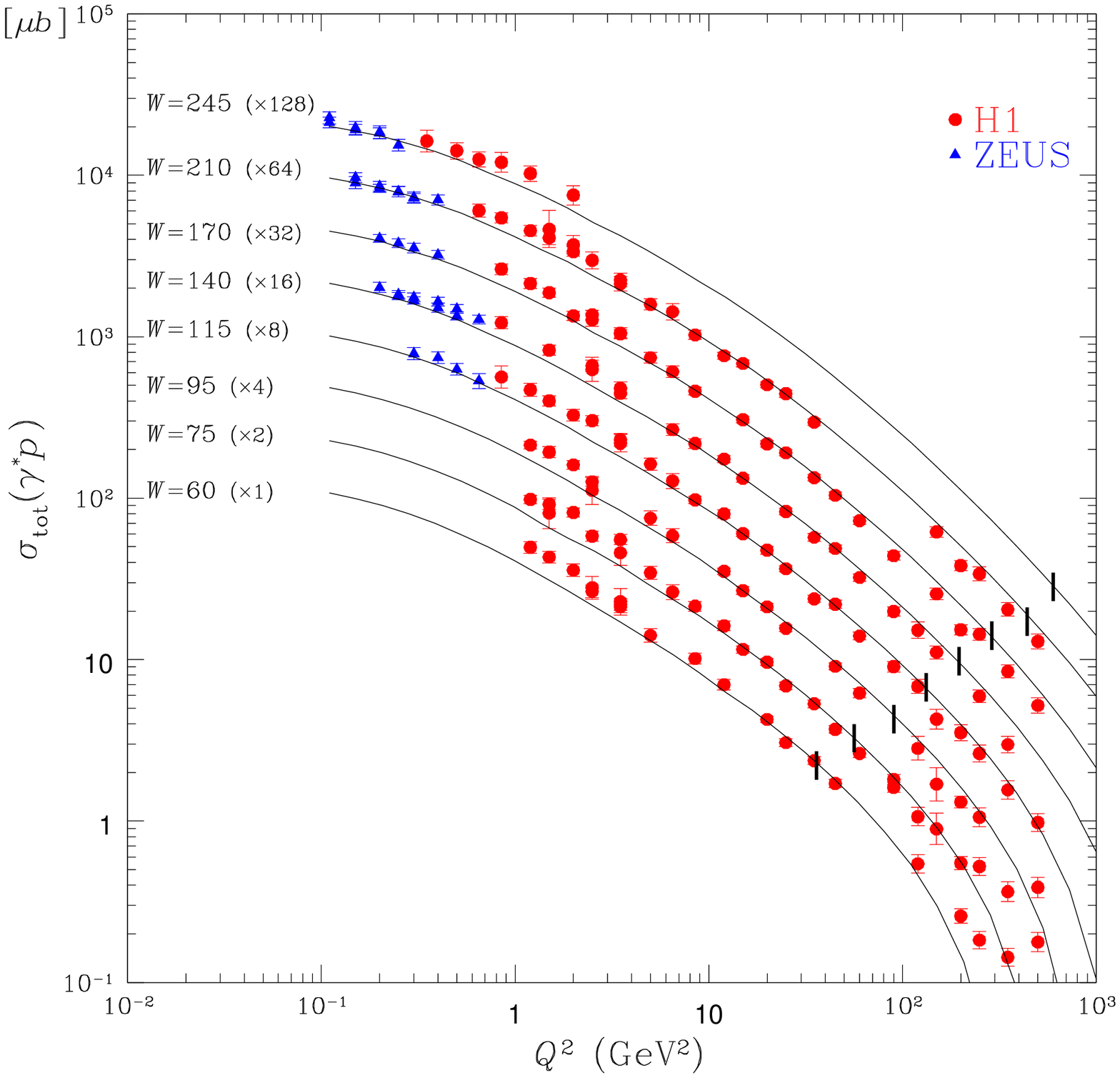,width=8.5cm}\\
\end{tabular}
\caption{\it $\sigma_{tot}(\gamma^* + p )$ in our model. From these data
we extracted  $Q^2_{0T} = 0.7 \div 0.9\,\,GeV^2$ and $Q^2_{0L} <0.4
\,\,GeV^2$. The vertical line corresponds to $x = 0.01$.}
\label{sgma}
 \end{figure}

As has been discussed the matching between ``soft" and ``hard" processes
is not very sensitive to the saturation scale since we will show in the
next section that the shadowing corrections to $F_2$ are rather small
\cite{AGL}. Therefore, we use $\sigma_{tot} (\gamma^* p )$ 
  to extract  the separation scale from the experimental data
\cite{HERADATA,HERAREV}. It turns
out that
$r^{sep}_{\perp} \approx 1/Q_{0T} \approx
1/(0.85 \div 0.95 )\,GeV^{-1}$ for the  transverse polarized photon, while 
 $r^{sep}_{\perp} \approx 1/Q_{0L}   \geq 0.6\,GeV^{-1}$ \cite{GLMPHP}.
Fig.~\ref{sgma} shows our description of the experimental data at $x <
0.01$

\subsection{$\mathbf{F_2}$.}
\begin{figure}
\begin{center}
\epsfxsize=12cm
\leavevmode
\hbox{ \epsffile{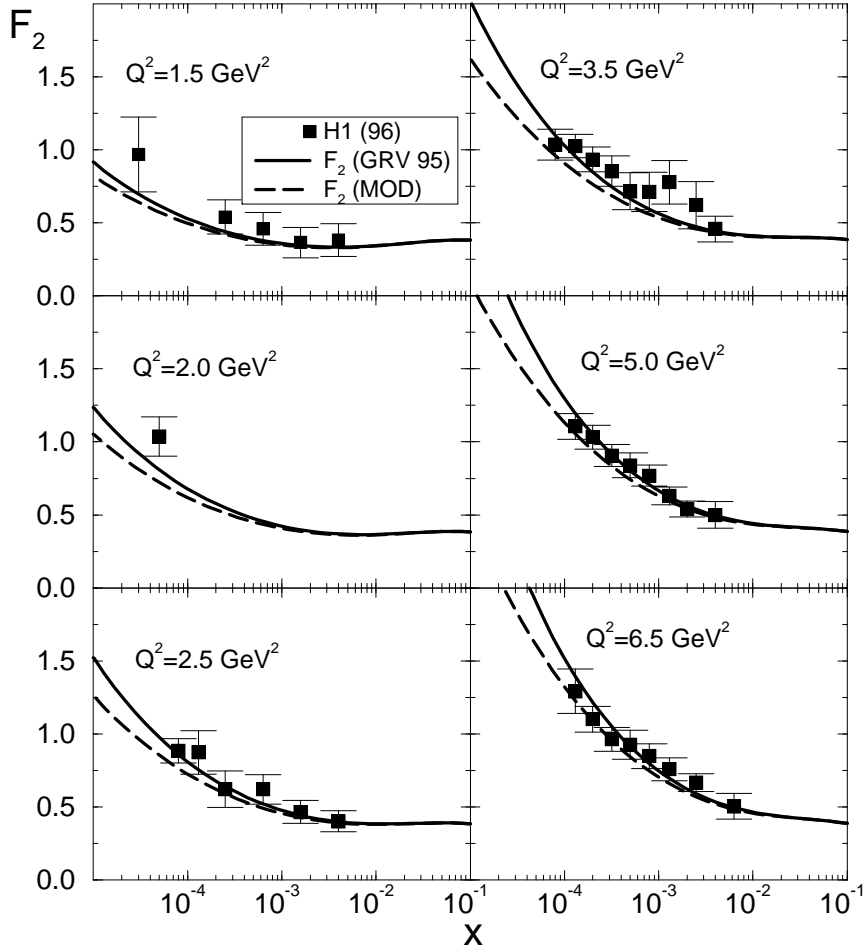}}
\end{center}
\caption{\it $F_2$ with  ( dashed line) and  without SC in our model.
SC has been taken into account for distances shorter than the separation
scale ($r_{\perp} < r^{sep}_{\perp} =  1/Q_0$).}
\label{f2}
\end{figure}

Our main conclusion that the global features of $F_2$ is not very
sensitive to the contributions of the SC from distances $r_{\perp}\, <\,
r^{sep}_{\perp}$ \cite{AGL}. In Fig.~ \ref{f2} we plot the result
of our estimates for low $Q^2 = 1.5 \div 6.5 \,GeV^2$, for higher $Q^2$
the shadowing corrections are even smaller.

\subsection{$\mathbf{xG(Q^2,x)}$}
Our calculations for $xG(x,Q^2) $ has been discussed in section V1 ( see
Figs. 6-a - 6-c and Ref. \cite{AGL} ).  One can see from these
figures that unlike  the case
of $F_2$,  the shadowing corrections are rather large and they 
tame the increase of the gluon density. Actually, all indications of 
high density QCD effects, that we see in the HERA data and will discuss
below,  are  due to 
strong screening in the gluon channel.

\subsection{ $\mathbf{F_2}$ Slope. }
\begin{figure}[hptb]
\begin{center}
\psfig{file=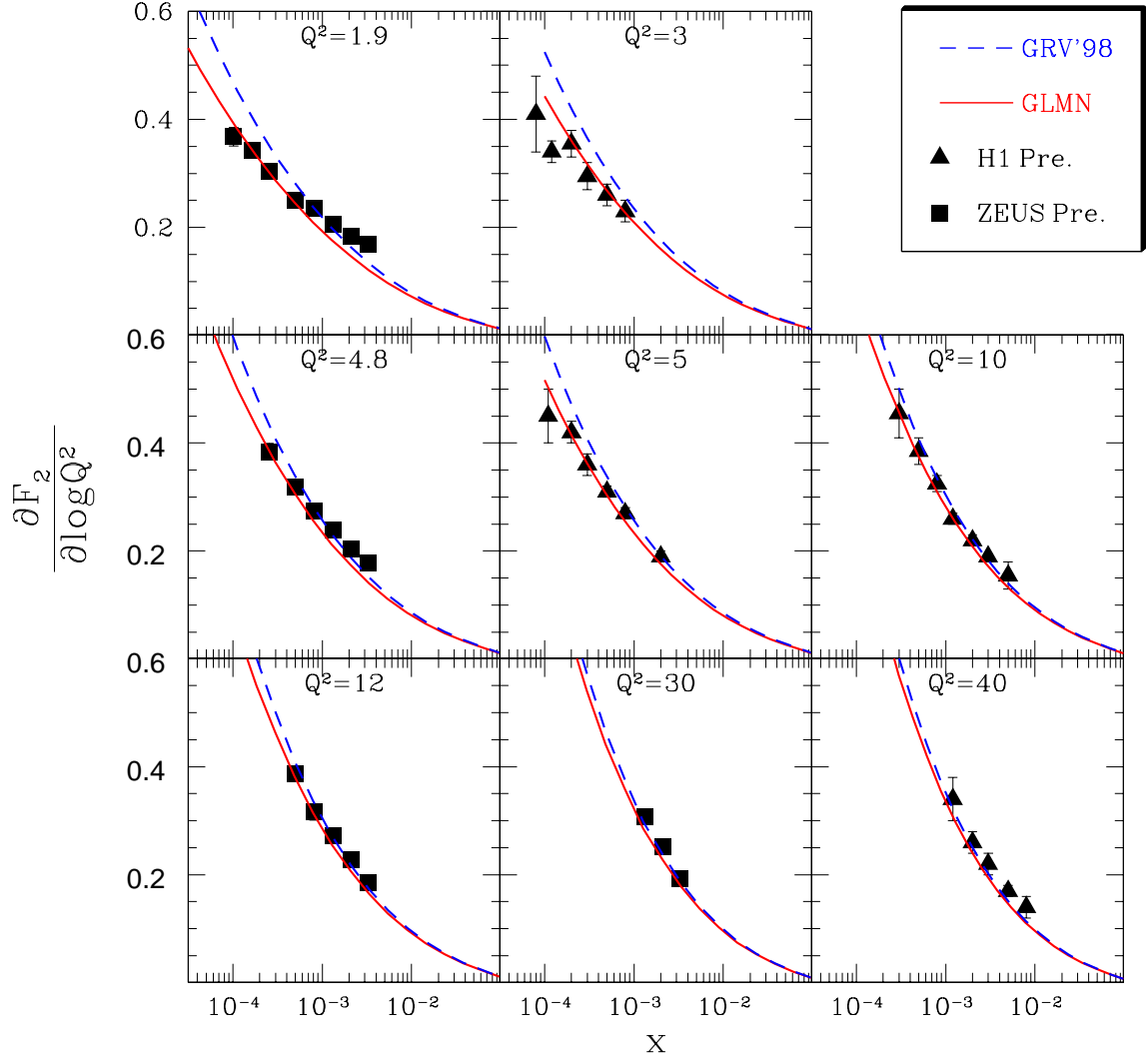,width=160mm}
\end{center}
\caption{\it $F_2$-slope in GRV'98  and in our model with
GRV'98
input for gluon structure function. Data are taken from figures of Ref.
\protect\cite{H1SLP} and from Ref. \protect\cite{ZEUSSLP}.}
\label{slp1}
\end{figure}

We consider the $Q^2$ behavior of the  $F_2$ slope, $ \frac{d
F_2}{d ln Q^2}$,
as the best experimental indication of
the
strong SC or other effects  of high density QCD. Our study
\cite{GLMSLP,OSAKA}  shows that none
of
available parameterization of the DGLAP equation can reproduce the
experimental data, while our model does this. More than that, 
two different parameterization such as GRV'94 and GRV'98 both describe the
data
well after taking into account SC in the framework of our model.
Fig.~\ref{slp1}   shows the comparison of the GRV'98 parameterization with
the 
H1 data. One can see that GRV'98 which was invented to describe ZEUS data
( Caldwell-plot) failed to fit H1 results but in our model we can fit H1
data
without any additional change of  parameters.

\begin{figure}
\begin{center}
\epsfxsize=13cm
\leavevmode
\hbox{ \epsffile{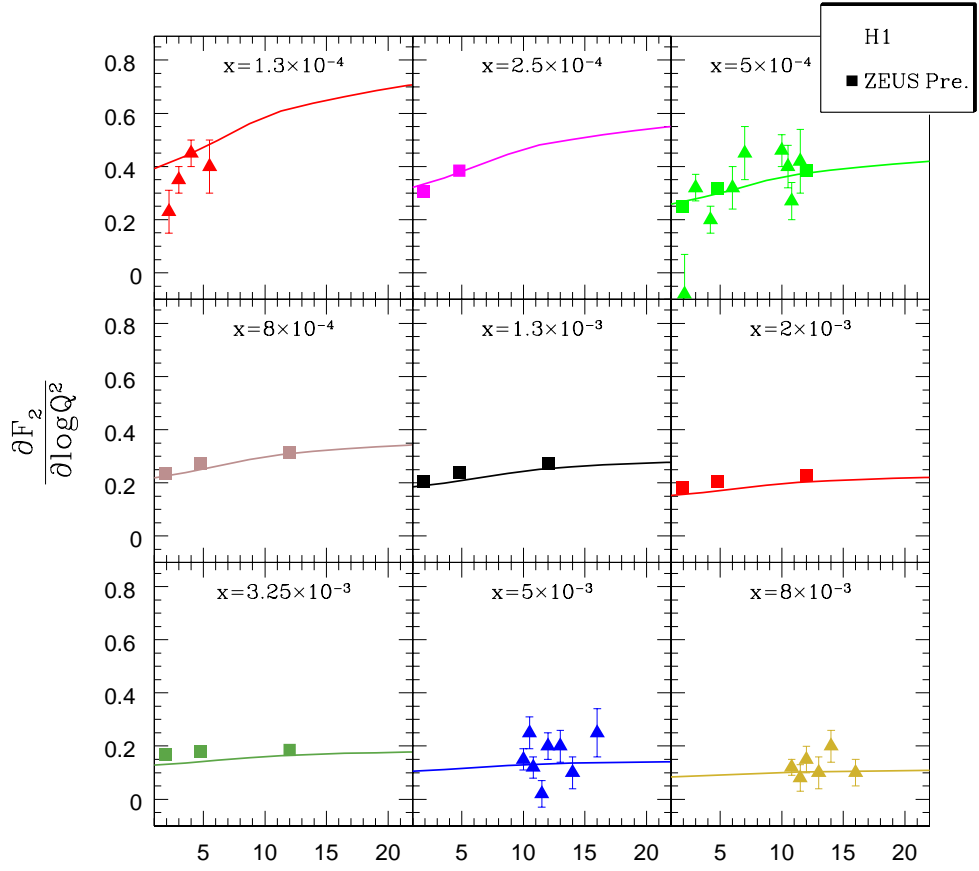}}
\end{center}
\caption{\it The $F_2$ slope at fixed $x$ versus $Q^2$.Data are taken from
figures of Ref.   
\protect\cite{H1SLP} and from Ref. \protect\cite{ZEUSSLP}.}
\label{slp2}
\end{figure}

Fig.~\ref{slp2}  presents   the `ideal' Caldwell plot from the point of
the gluon saturation (see section VII): the $F_2$ slope as function of   
$Q^2$ at fixed $x$.  One can see  two features in this figure: (i) our 
model describes the HERA data quite well; and (ii) the data  as
well as our calculation do not show
 any maximum which could be interpret as a
saturation scale.  Our model is also able to describe the $Q^2$ behaviour
of the $F_2$ at fixed $W$ which shows  maxima. It means that the maxima
in the 
fixed  $W$ plot does not reflect the saturation phase of the
parton
cascade but are an artifact of the particular choice of the kinematic
variable.

\begin{figure}
\begin{center}
\epsfxsize=13cm
\leavevmode
\hbox{ \epsffile{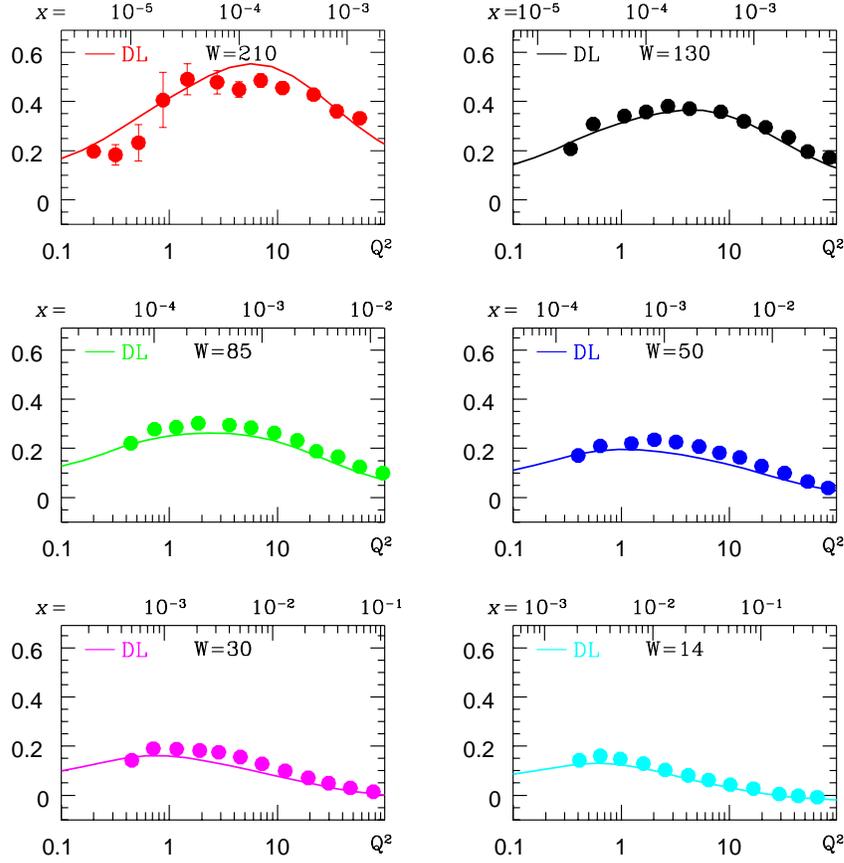}}
\end{center}
\caption{\it The $F_2$ slope at fixed $W$ versus $Q^2$. Curves are
calculated In Donnachie-Landshoff model \protect\cite{DL2P}. Data are taken
from figures of Ref.
\protect\cite{H1SLP} and from Ref. \protect\cite{ZEUSSLP}.}
\label{slp3}
\end{figure}

 We have also studied an alternative approach, namely, the
Donnachie-Landshoff model for matching the ``soft" and
 the ``hard" interactions
\cite{DL2P}: the sum of ``soft" and ``hard" Pomerons (see
Fig.~\ref{slp3}). One can see that this model is as successful in
 describing the data as is
 our model. This  means that at the
moment we cannot claim that the
data show a  saturation effect. We have a much more modest claim:
{\it The current data can be described in two different approaches, either
due to  gluon saturation or due to matching between ``soft" and ``hard" 
contributions ( ``soft" and ``hard" Pomerons ) but at rather large momenta
( about $1 - 2 \,GeV$ ).}  In spite of the fact that we failed to arrive
at  a
unique conclusion one can see that measuring the $F_2$ slope  gives new
 and interesting information about QCD  dynamics.

\subsection{Energy dependence of the inclusive diffraction cross section  
$\sigma^{diff}_{tot}$ and  the ratio
$\mathbf{\frac{\sigma_{diff}}{\sigma_{tot}}}$.}

In a saturation approach the typical distances that are dominant in the
diffractive production, are of the order of the saturation scale
$r^{saturation}_{\perp}$ \cite{SAT,GLMINDD,LW}, unlike the total
cross
section where they are of the order of $(1/Q)$.n One can see this 
directly from Fig.~\ref{pcsd} which shows that in the saturation region a
hadron appears  as  a  diffraction grid of size
$r^{saturation}_{\perp}$. This fact is in a perfect agreement with the
HERA experimental data \cite{HERADD} as well as with our estimates
\cite{GLMINDD}.

However,  we cannot reproduce in our model\cite{GLMRDT}
the experimentally observed fact that the
ratio  $ \sigma^{diff}_{tot}/\sigma_{tot}$ is almost energy independent.
 The  dependence on energy in our calculations stems partly 
 from distances shorter than $0.3 fm $ (see Ref. \cite{GLMRDT}
for details ) where we can trust the
DGLAP evolution equations. The DGLAP evolution  was included in our model
in contrast to 
the Golec-Biernat and Wusthoff model \cite{GW} which includes the
saturation scale
but  has a very oversimplified behaviour at short $r_{\perp}$ ($
\sigma_{dipole} \propto r^2_{\perp} $ ).

\begin{figure}[htbp]
\begin{tabular}{c c}
 $\mathbf{Q^2 = 8 \,GeV^2 }$ & $\mathbf{Q^2 = 14 \,GeV^2 }$ \\
  \psfig{file=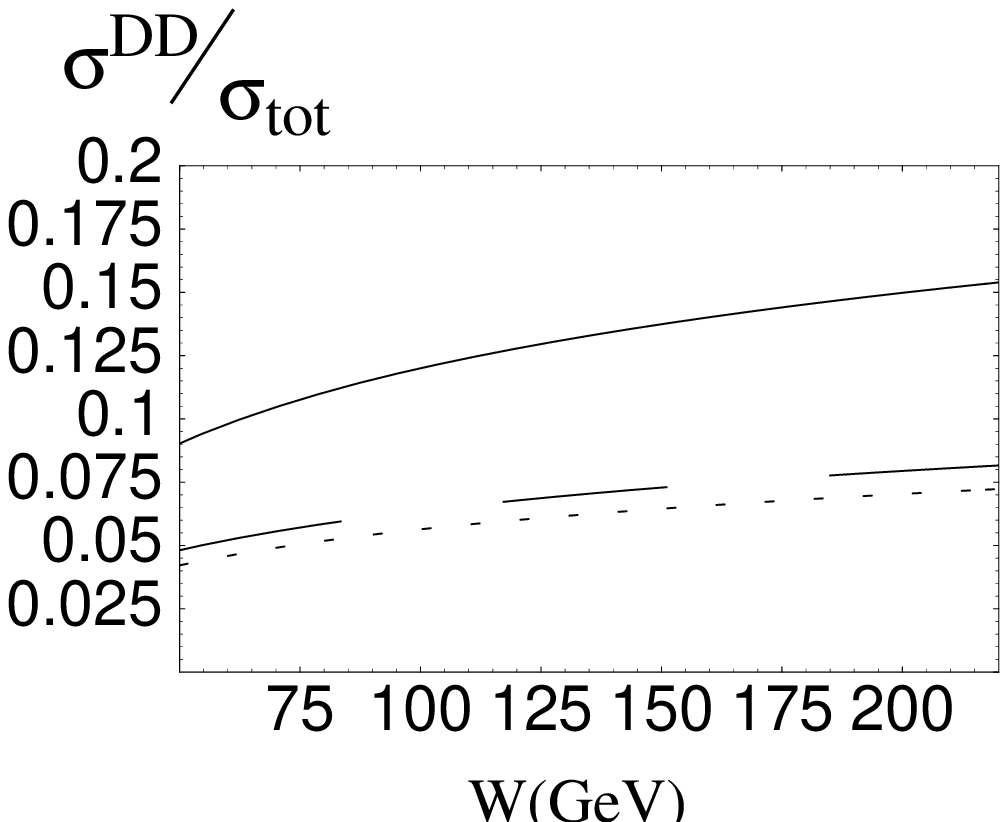,width=80mm,height=60mm}&
\psfig{file=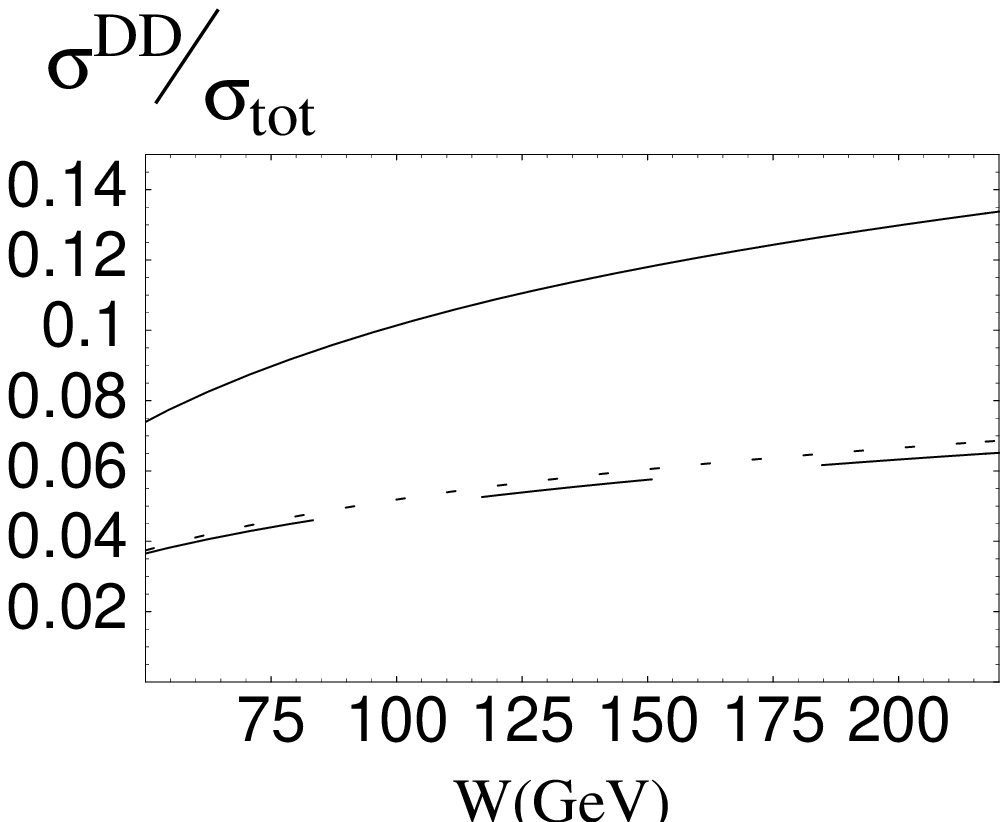,width=80mm,height=60mm}\\
$\mathbf{Q^2 = 27 \,GeV^2 }$ & $\mathbf{Q^2 = 60 \,GeV^2 }$ \\
 \psfig{file=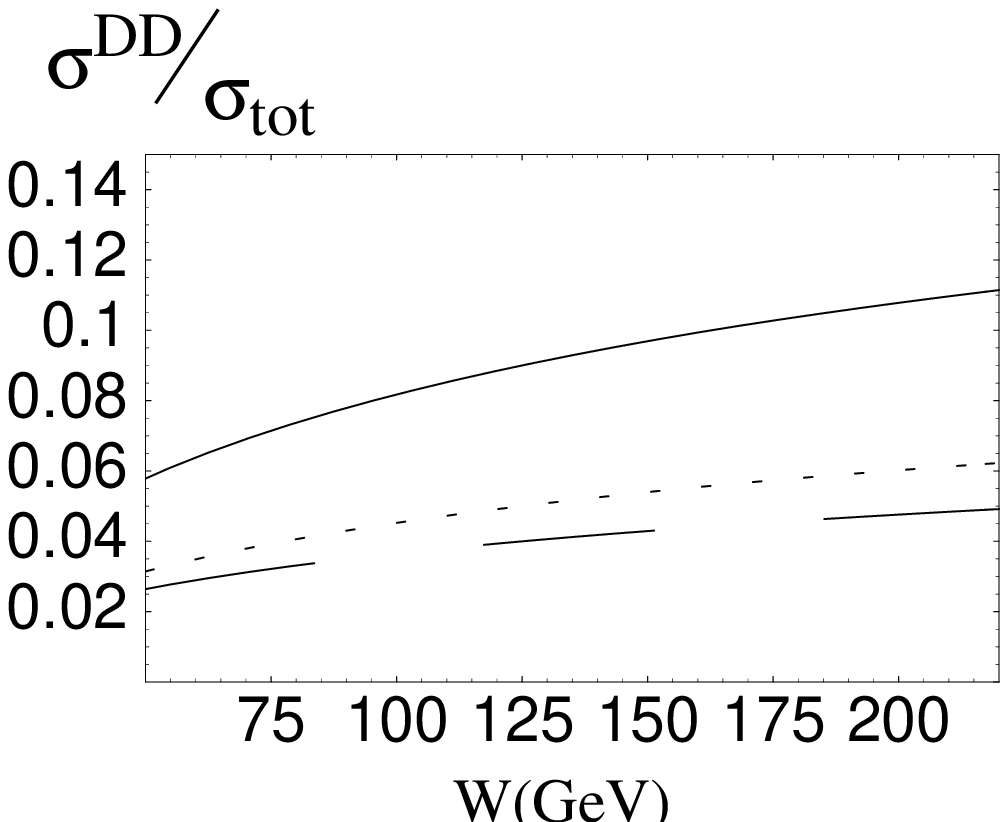,width=80mm,height=60mm}&
\psfig{file=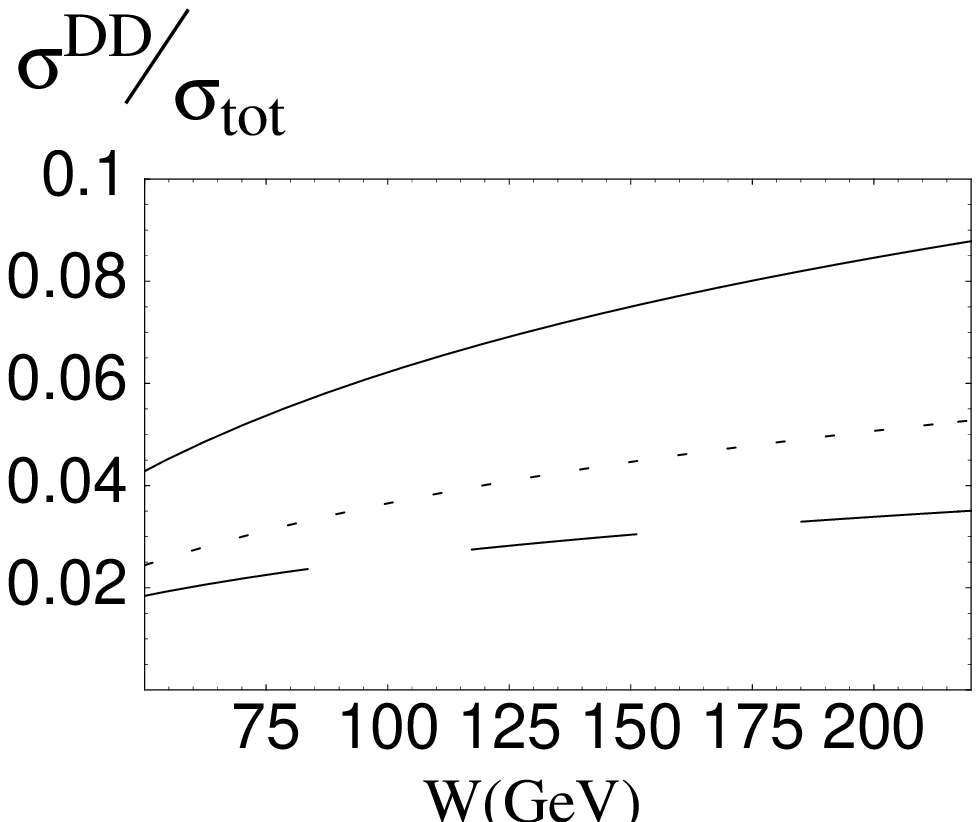,width=80mm,height=60mm}\\
\end{tabular}
\caption{ \it $\sigma^{diff}/sigma_{tot}$ vesus energy in our model. Solid
line is 
total contribution, dashed line is 
   $q \bar q $  and  dotted line is  $q \bar q + G  $  contributions.}
\end{figure}

\subsection{ Energy dependence of J/$\mathbf{\Psi}$ production.}
\begin{figure}
\begin{center}
\psfig{file=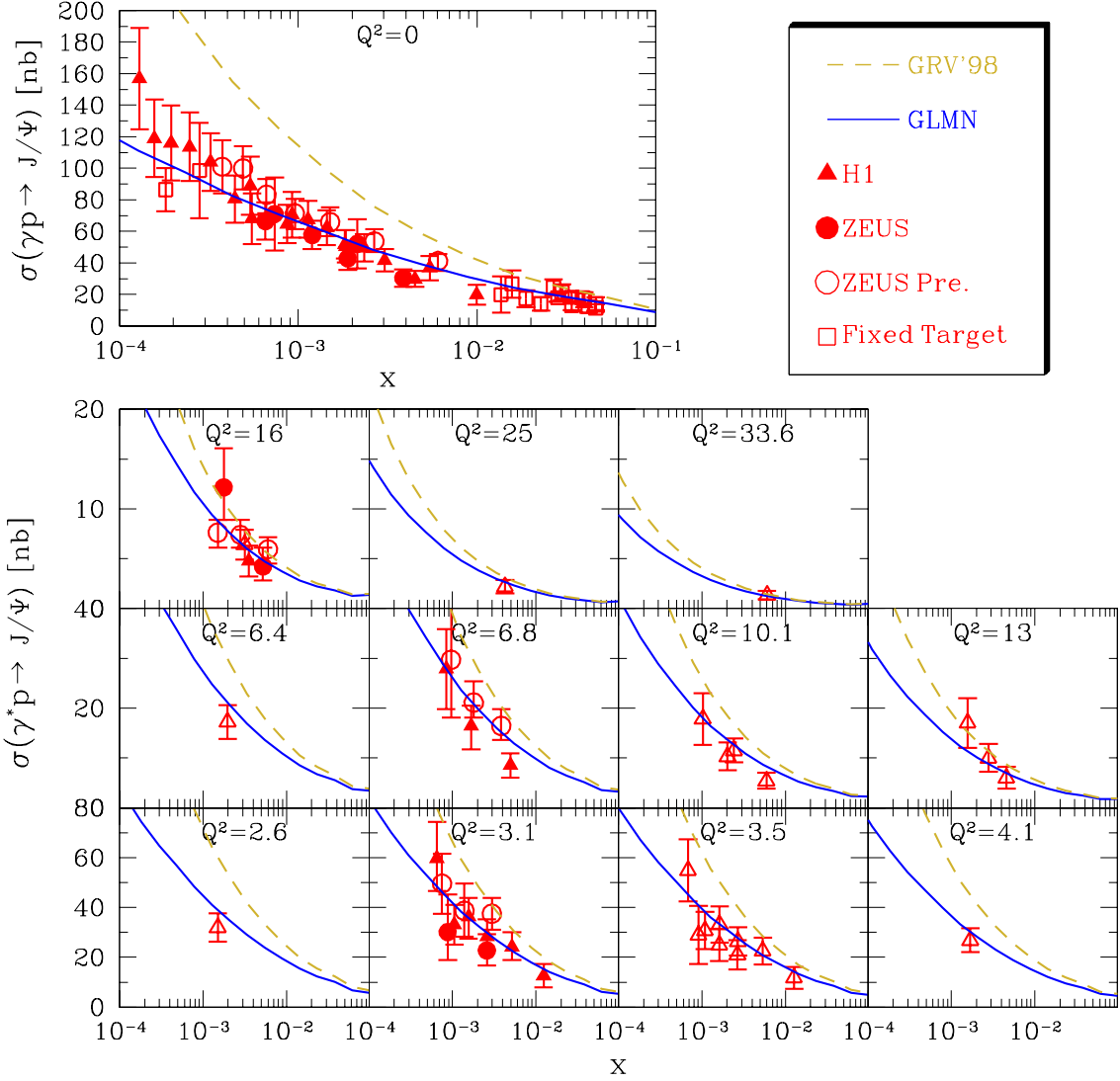,width=16cm} 
\end{center}
\caption{\it Examples of the description of J/$\Psi$ DIS and photo
production. We got excellent n.d.f. $ <$ 1.}
\label{psi}
 \end{figure}

We claim that the energy behaviour of the exclusive vector meson
production, especially, J/$\Psi$-production is very discriminating with
respect to SC (see Fig. ~\ref{psi}). In spite of large errors related to
uncertainties due to our
poor knowledge of the wave function of vector mesons, this uncertainty 
contributes mostly to the normalization  of the cross section while the
energy
slope is still a source of the information on the SC. We believe that
at the moment there is no parameterization of the DGLAP evolution equation
which is able to describe simalteneously the $F_2$-slope and
the energy dependence of the J/$\Psi$ photo production where a  huge
amount of new data has been accumulated ( see Ref. \cite{OSAKA} and
references therein for more details ).

\subsection{ $\mathbf{t}$ - behaviour of hard diffractive leptoproduction
of vector meson.}

\begin{figure}
\begin{tabular} {l l}
\psfig{file=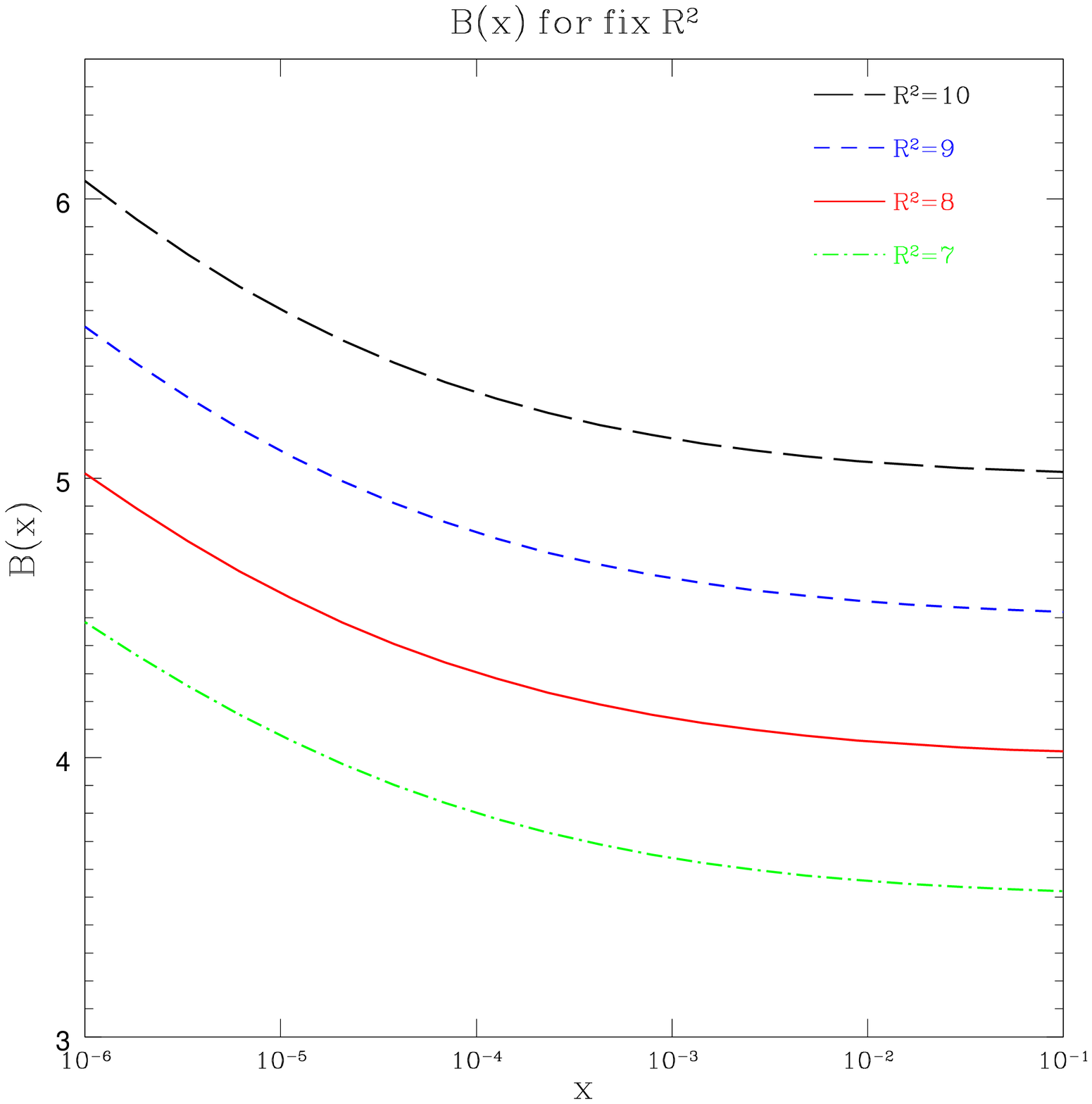,width=8.5cm} &
\psfig{file=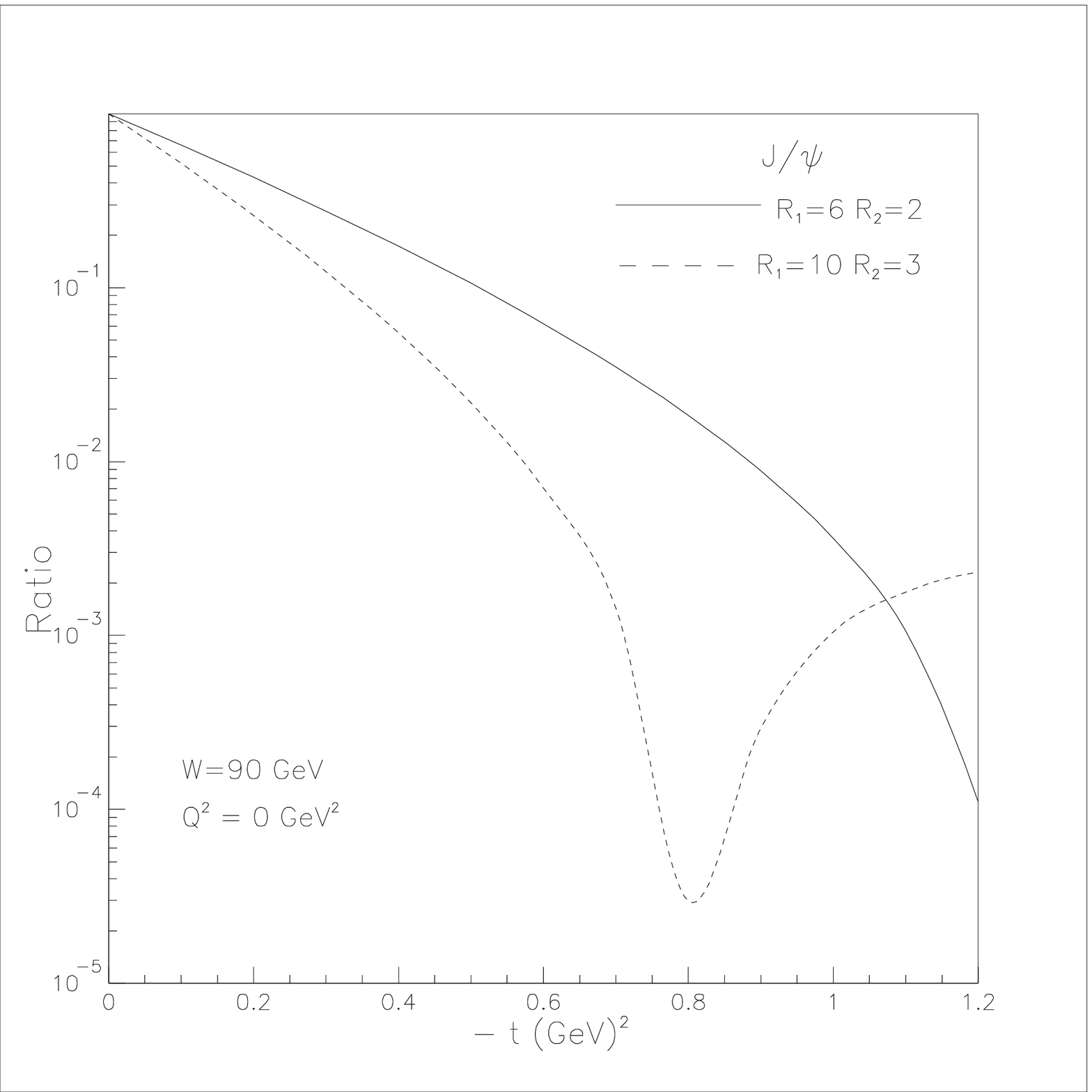,width=8.5cm}\\
 & \\
fig.15-a & Fig. 15-b \\
\end{tabular}
\caption{ \it The energy behaviour of the diffraction slope
 ( FIg. 15-a) and the $t$ - dependence (Fig. 15-b)  for J/$\Psi$
photo production. Fig.15-b corresponds to the two radii model for the
impact parameter dependence \protect\cite{GLMVP} of the profile function
which provides a better description of the experimental
data. Fig. 15-a is plotted for one
radius model. The experimental data are taken from
\protect\cite{PSIDATA}.}
\label{dslp}
\end{figure}

As has been mentioned, in our model we generate  a shrinkage of the
diffraction peak (see Fig.~\ref{dslp} ) as well as a damping of the energy
dependence. We want to draw the readers
attention to the fact that we expect a characteristic $t$-dependence with
possible minimum around $|t| = 0.8 \div 1.2 \,\,GeV^2$.

\section{Searching for new observables:}
\subsection{ $\mathbf{\sigma(\gamma  \,\,\gamma^*)}$.}
\begin{figure}
\begin{tabular} {l l}
\psfig{file=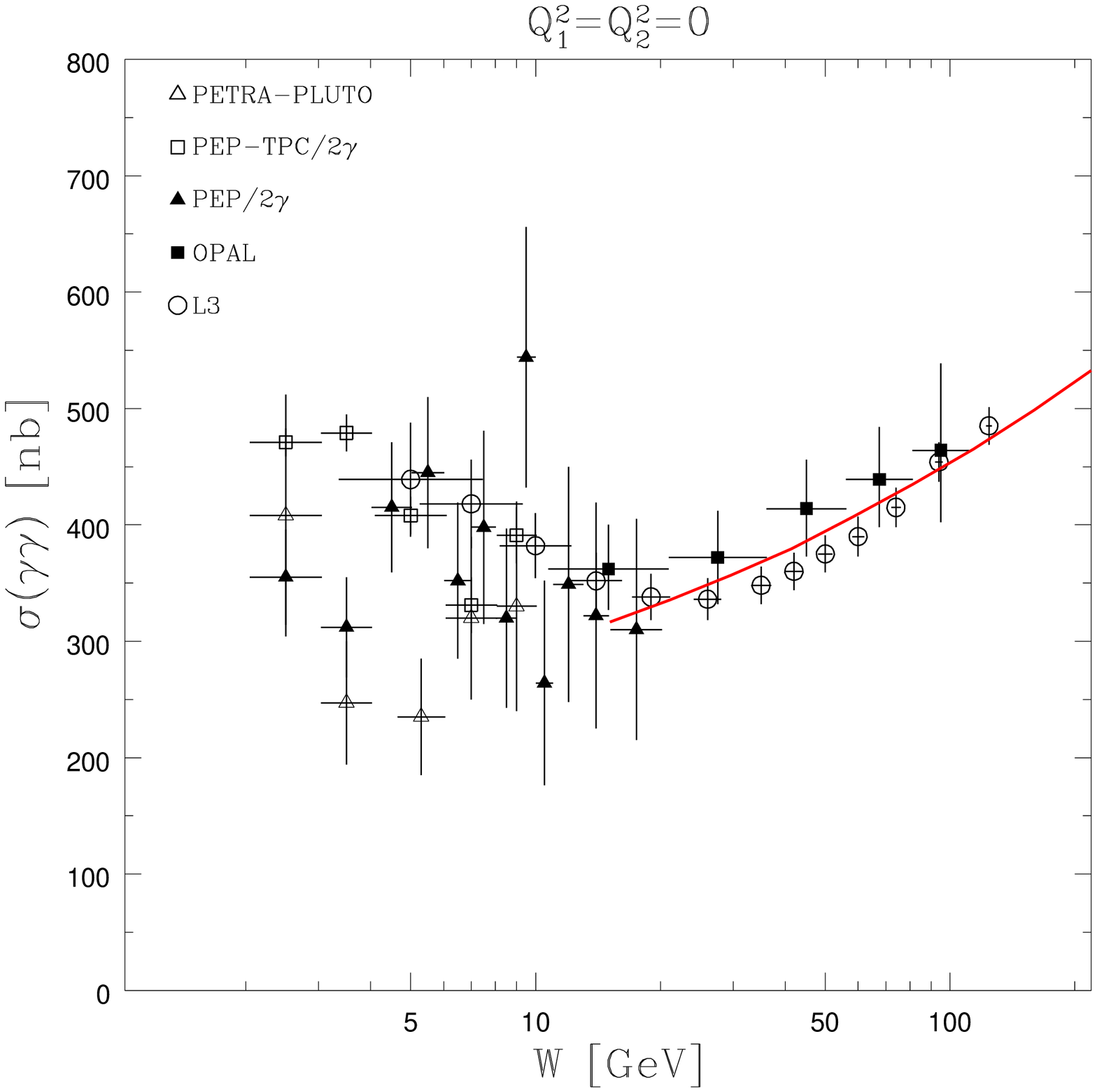,width=8.5cm} &
\psfig{file=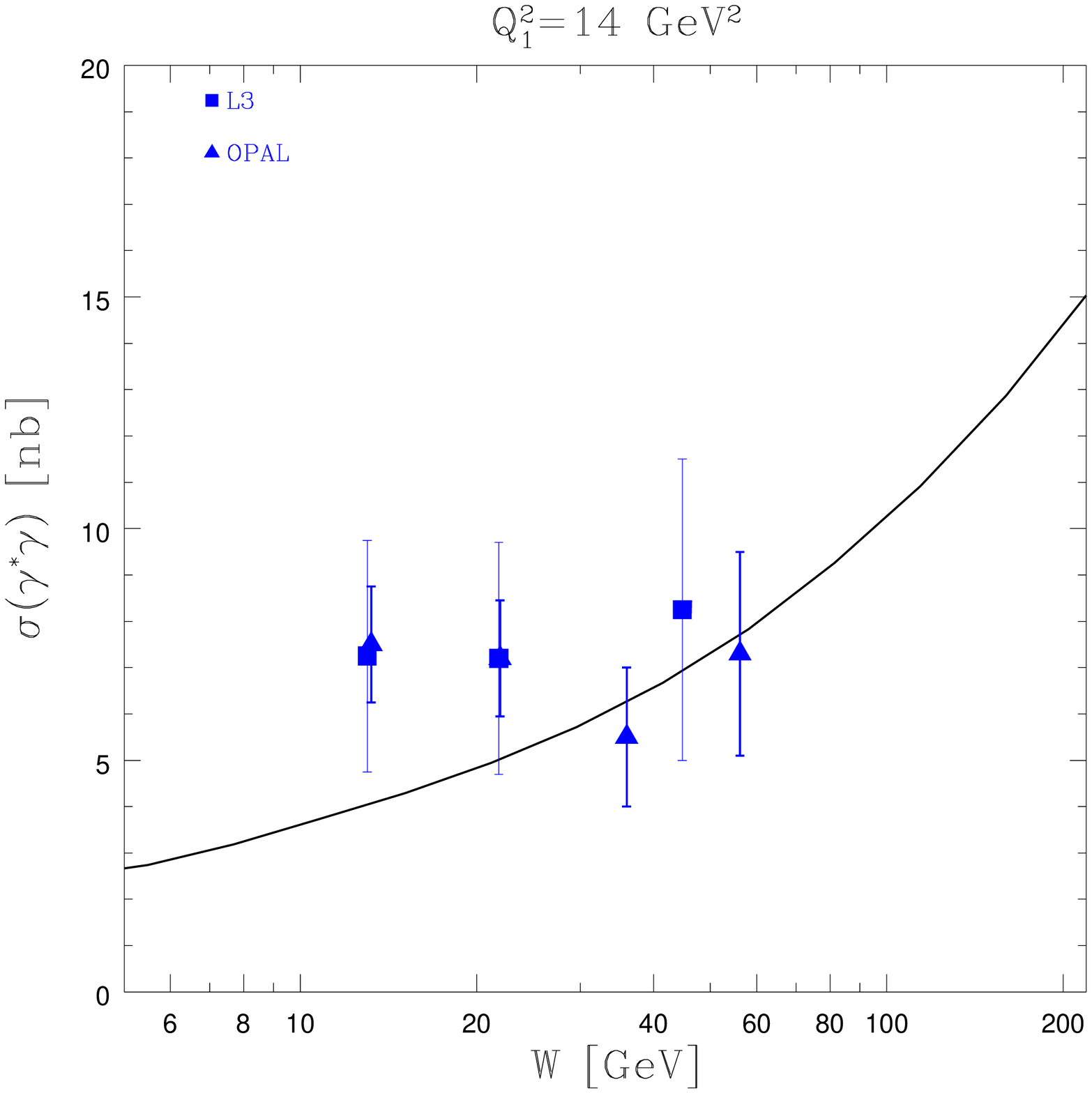,width=8.5cm}\\
\end{tabular}
\caption{\it  The description of $\gamma^* \,\gamma^*$ total cross section
at
different values of the photon vituality in our model. The experimental
data are taken
from Ref. \protect\cite{PHPHDATA}.}
\label{phph}
\end{figure}

The interaction of two virtual photons provides a unique opportunity to
measure
both the property of the BFKL Pomeron \cite{TPHPH} as well as to explore
the saturation region. By now we have checked that our model is able to
describe\cite{GLMPHPH} current experimental data for the
$\gamma \gamma^*$
total cross
section\cite{PHPHDATA} with the same parameters that we used for DIS with
the nucleon target (see Fig. ~\ref{phph} ). However, we cannot reproduce
in our approach the experimental data for the cross section of two virtual
photons with $Q^2_1\, \approx Q^2_2\, > 1\, GeV^2$. Perhaps, we need to
improve our approach including the BFKL contribution.

\subsection{Maxima in ratios.}

\begin{figure}
\begin{tabular}{l l }
\psfig{file=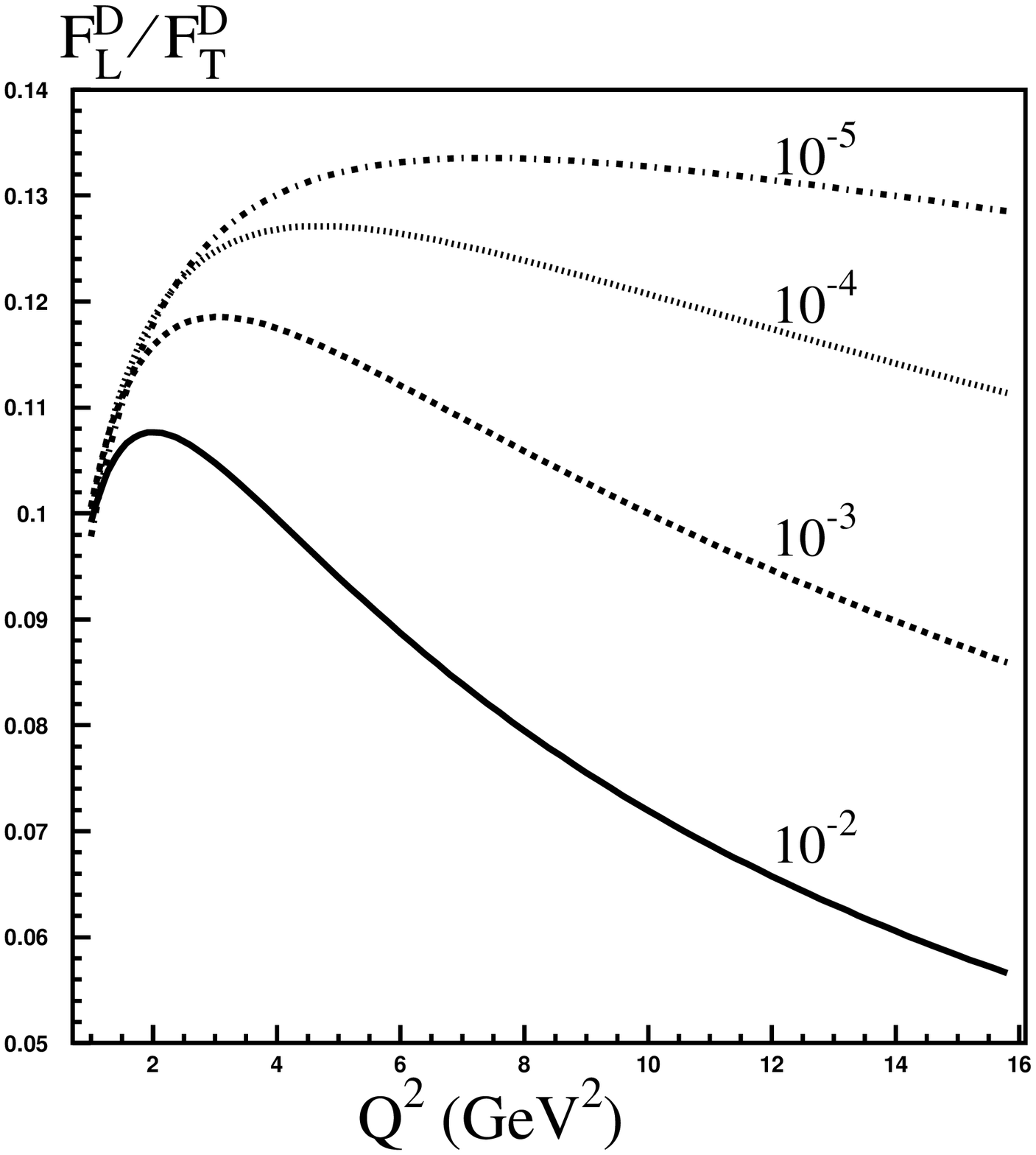,width=8.5cm} &
\psfig{file=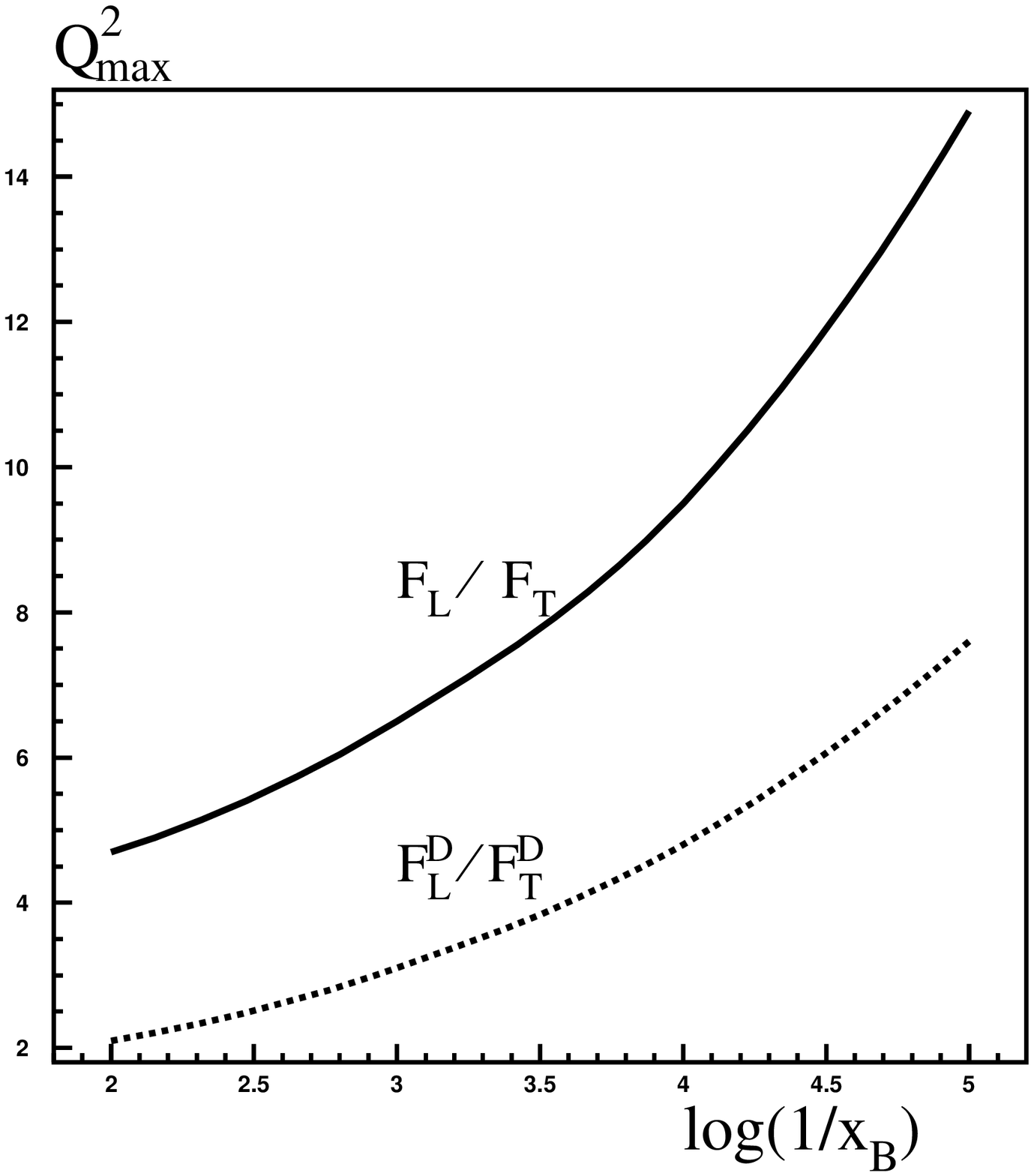,width=8.5cm}\\
\end{tabular}
\caption{\it The ratio of $F^D_L/F^D_T$ as function of $Q^2$ at different
values of $x$ and the behavior of $Q^2_{max}(x)$ as function of $x$.}
\label{max}
 \end{figure}

In our attempt to find an improved  observable which will be more
sensitive to
the saturation scale we study the $Q^2$ behaviour of the ratios:
$F_L/F_T$ and $F^D_L/F^D_T$ for longitudinal and transverse structure
function for inclusive DIS and for diffraction in DIS\cite{GLMHT}. We
found that these ratios have maxima at $Q^2 = Q^2_{max}(x)$  which are
moving as a function of $x$. In Fig.~\ref{max} we have plotted some
examples of
these ratios and the behaviour of $Q_{max}(x)$. It appears that
$Q_{max}(x)$ is  a  simple function of the saturation scale
$Q^2_s(x)$ of \eq{EQ2}.

~

\subsection{Higher twist contributions.}
As we have mentioned, one of the attractive features of our model is the
fact
that  we are able to describe the higher twist contribution in
accordance with all known theoretical information. Our calculations
confirm the result of Ref.\cite{HTM} that there is an almost full
cancellation of higher twist
contribution in $F_2$ while they give  substantial contributions
separately in $F_L$ and $F_T$ as well as in $F^D$.  We hope, that the
twist analysis will help us to find a scale when the higher twist
contribution will be of the order of the leading one. It is well known
\cite{SAT}, that in a saturation approach at $Q^2 = Q^2_s(x)$ all twist
contributions are of the same order. Therefore, our twist analysis will
assist in finding a kinematic range of $Q^2$ and $x$ where the higher
twist
should be important and where we expect a manifestation of the gluon
saturation.

\section{Our plans for the near future.}
 We list here the problems that we have started to work on:
\begin{enumerate}
\item\,\,\, Large $p_t$-jet  production in photo production. The first
calculation indicates a maximum at the saturation scale in $p_t$
-spectrum;
\item\,\,\,Heavy quark production and especially open charm production
will appear soon;
\item\,\,\,Using our model we intend to calculate the inclusive
observables and to examine their sensitivity to  SC;
\item\,\,\,We have started, and we  are going to continue,  to check our
approach
with the $\gamma^* \gamma^*$- data on total cross section;
\item\,\,\,The calculations for gluon structure function as well as for
$F_2$ for nuclear targets has been performed, as well as predictions for
the
total diffractive cross sections for DIS with nuclei. We have
extended our twist analysis to DIS with nucleus\cite{GLMA}  to see whether
we will be
able to separate the higher twist contribution using A-dependence. We have
started  to calculate all above mentioned  observables for DIS with
a nuclear
target, hoping that the future experiment for DIS with nuclei will open a
new
dimension in our investigation of the saturation phenomenon and will be
complementary to HERA data.

 \end{enumerate}

\section{Feedback for theory:}

It should be mentioned that our model or a similar type of approach is
needed to solve the nonlinear evolution equation that have been discussed
in section IV. The essential requirements to solve 
\eq{GLRINT} are:
\begin{itemize}
\item \quad  a correct initial distribution, which should be 
 of the eikonal type
as was shown by Kovchegov \cite{EQ};
\item \quad values of the parameters that enter \eq{GLRINT}, such as the
$b_t$-distribution and a separation scale;
\item \quad a relationship between the dipole scattering amplitude
$a^{el}(r^2_{\perp},b_t;x)$ and the gluon structure function.
\end{itemize}

Our model provides an answer to all formulated questions and can be used
as
an  initial distribution for the nonlinear evolution equation (see
\eq{GLRINT} ).  Indeed, our model gives the eikonal description of the
experimental data which is suited to be used as the initial condition for
\eq{GLRINT}. The only item that we  need to add is the relation
between the gluon structure function and the dipole scattering amplitude. 
This relation is a natural generalization of \eq{GM3}, namely,

\beq \label{RE1}
x G(x, \frac{4}{r^2_{\perp}})\,\,\,\,=\,\,\,\,\frac{4 R^2}{\pi}\,\int
\,\,d^2 b_t \,\int^1_{x}\,\frac{d
x'}{x'}\,\int^{\infty}_{r_{\perp}}\,\,\frac{ d
r^2_{\perp}}{r^4_{\perp}}\,\,2\,\,Im \,\,a (r^2_{\perp},b_t; x)\,\,.
\eeq

A similar relation for the $F_2$ structure function stems directly from
the
relationship between the measured cross section ( $ \sigma(\gamma^* p )$
) and $F_2$  ( $ \sigma(\gamma^* p )  = \frac{4 \pi^2}{Q^2} F_2$ ) and
from
\eq{SDF} and \eq{GM1}. We assumed that for a gluon probe \cite{DOF3} we
have the same relation between the gluon structure function and the
measured cross section. It should be stressed that \eq{RE1} coincides with
the definition of the gluon structure function in the DGLAP limit.

Based on the  model developed for the initial condition we will
attempt
to solve the evolution equation ( see \eq{GLRINT}) and so
provide  reliable estimates for the structure functions in the LHC
kinematic region.

\section{Comparison with other models:}

the models that are on the market  fall into two classes: (i)
 models that introduce the saturation scale in addition to the separation
one ( see our model and  Refs. \cite{GW,MCD,MSF} ), and (ii)
models that  only have a separation scale (see
Refs.\cite{DL,DL2P,MCD,MSHA}).  We discuss here only models
that reproduce the same data or almost the same data as ours with a chi
squared of approximately the same value. There are two such 
models\footnote{Model of
Ref. \cite{MSF} also introduces a saturation scale but in a quite
different way than in our model and in Golec-Biernat and Wusthoff model
\cite{GW}. However, this model is only in a very premature stage of
development and has not described the HERA experimental data with a good
chi squared. One can find the comparison of this model with other
competing models
in Refs. 
\cite{MCD,GW,MSHA}.}

Our model and Golec-Biernat and Wusthoff model \cite{GW} are in the first
class: both of them introduce a saturation scale and we will discuss the
difference between them a little  later. The Forshaw, Kelley and Shaw
model\cite{MSHA}
suggested a different approach based on ideas of Donnachie and Landshoff
model \cite{DL,DL2P}. They use \eq{SDF}, and for the dipole cross section 
they suggest the sum of ``soft" and ``hard" contribution:
\beq \label{MSHA1}
\sigma_{dipole}(r^2_{\perp},W)
\,\,=\,\,f_S(r_{\perp})\,W^{\lambda_S}\,\,+\,\,f_H(r_{\perp}\,\,
(r^2_{\perp}\,\,W^2)^{\lambda_H}\\,
\eeq
with the following limits of 
\be
f_S(r_{\perp})\,\,\,\, & \longrightarrow &\,\,\,\,Const \,\,\,
at\,\,\,\,r_{\perp} \,\geq\,0.8 \div\,1\,fm \,\,;\label{MSHA2}\\
f_H(r_{\perp})\,\,\,\, & \longrightarrow  &\,\,\,\,r^2_{\perp}\,\,\,
at\,\,\,\,r_{\perp} \,\leq\,028 \div\,0.3\,fm\,\,. \label{MSHA3}
\ee
Therefore, in \eq{MSHA1} one can recognize the ``soft" Pomeron in the
first term, while the second one at short distances (  $r_{\perp}
\ll r^{sep}_{\perp}$ ) depends on $x$ as it should  for the ``hard"
contribution.

The success of this model shows us that the  HERA data alone cannot
distinguish between a rather small separation scale and  saturation of the
gluon density. In this model ``hard" and ``soft" are compatible at
$r_{\perp} \approx
0.4 \,fm $ which is rather small.

There is no difference in principle between the Golec-Wusthoff model and
ours
since both of them use the same physical picture that follows from the
high density QCD. However, in our model the saturation scale can be
calculated (see \eq{EQ2} ) and all our results have the correct DGLAP
limit at
short distances. On the other  hand,  Golec-Wusthoff model introduces the
saturation scale as a phenomenological parameter, has an extremely simple
form and reproduces the HERA data on the energy behaviour of ratio for
$\sigma^{diff}_{tot}/\sigma_{tot}$.  In other words,  Golec-Wusthoff model
investigates the extreme limit: { \it all physics observed at HERA is
related to
a
saturation scale}. It is very encouraging that such a rudimentary idea
does not
contradict all experimental data from HERA.

\begin{figure}
\begin{tabular}{c c}
\psfig{file=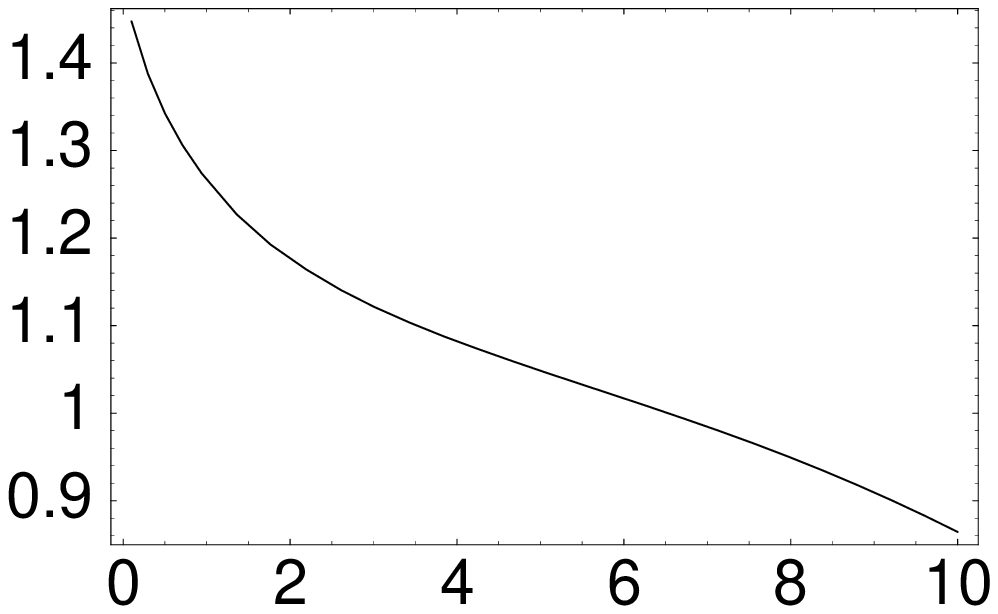,width=7.5cm} &
\psfig{file=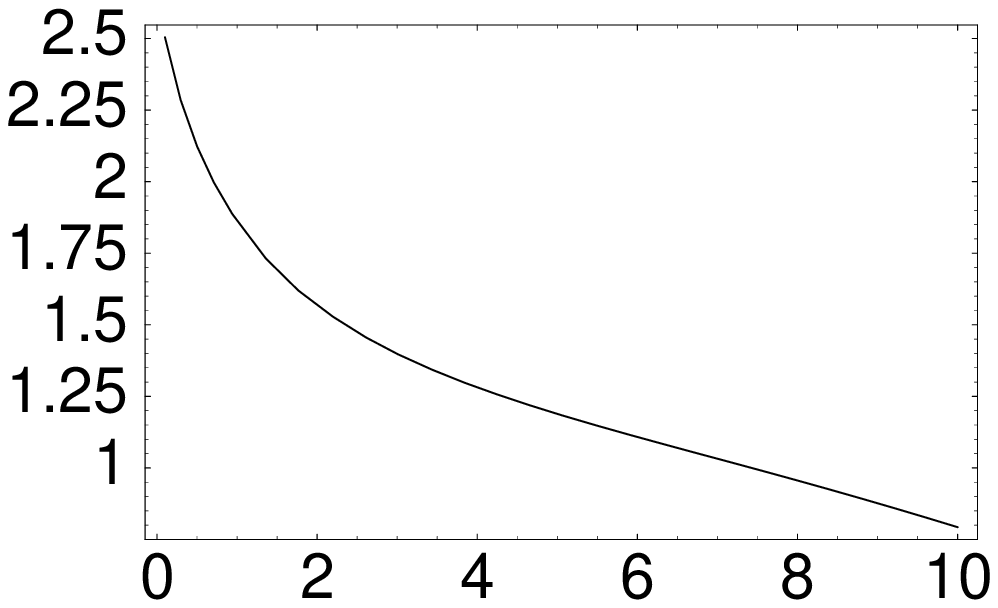,width=7.5cm}\\
       & \\
Fig. 18-a & Fig.18-b\
\end{tabular}
\caption{\it The ratio of our dipole cross section to Golec-Biernat and
Wusthoff dipole cross section at $x = 10^{-3}$ versus $r^2_{\perp} (
GeV^{-2})$ for total cross section ( Fig.18-a) and for diffractive cross
section (Fig.18-b).}
\label{comp}
\end{figure}

In Fig.~\ref{comp} we plot the ratio of our dipole cross section to
Golec-Wusthoff one. More  work is needed to understand how much
of the disagreement stems from attractive features of our models such as
correct matching with the DGLAP evolution equation, and how much of them 
can be explained by the fact that our model underestimates the value of SC
at rather low $Q^2 \approx 1\div 2 \,GeV^2$.

It is incorrect to say that it  is not possibile to distinguish one
model from another. For example, for diffractive production it is
important to take into account the diffractive dissociation in $ q \bar q
G$ channel. This channel can be considered as the production of a dipole
but
with a charge which is twice larger ( for $N_c > 1 $ )  than for a
$ q \bar q$ dipole. In spite of this fact in the two models \cite{GW,MSHA} 
the dipole for $ q \bar q G$ - channel was treated in the same way as $ q
\bar q$ dipole ( in MFGS model\cite{MSF} it has not yet been considered
). 
 In our model which is based on the correct matching with the pQCD result,
the
propagation of $q \bar q G$ - dipole is quite different than for a
quark-anti quark
pair. This means that these models should have different predictions for
jet
production in the diffractive dissociation processes.

\section{Our answer:}
In our opinion HERA has reached a high density QCD domain because: (i)
HERA data display  a large value of the gluon structure function; (ii) no
contradictions with the asymptotic predictions of high density QCD have
been observed; and (iii) the numerical estimates of our model give a
natural description of the size of the deviation from the usual DGLAP
explanation. However, we  know the main shortcoming of our statement,
which is that  most
of the 
observed indications of high density effects can be described
alternatively
without assuming a saturation scale, but by matching ``soft" and ``hard"
processes at rather short distances. We want to stress three
aspects: 
\begin{enumerate}
\item \quad 
 The alternative approaches cannot describe or, perhaps, have
not described all data and, in particular, the energy dependence of the
J/$\Psi$ photon and DIS production \cite{OSAKA}. We found that a
simultaneous  analysis of this reaction and of the $F_2$ slope cannot be
done
without employing shadowing corrections;
\item \quad The gluon saturation is not an additional postulate of  
pQCD,
but follows from the QCD evolution equations in the high parton density
kinematic region, and our model describes the solution to these 
equations;
\item \quad The gluon saturation leads to the simplest Golec-Biernat and
Wusthoff  model which gives
an impressive description of the data.
\end{enumerate}

The fact that one can describe the HERA data without assuming  gluon
saturation is not very surprising as the saturation scale in the
HERA kinematic region is $ Q^2_s(x) = 1 \div 3 GeV^2$ while the typical
scale for the ``soft" Pomeron is also not very small and can be as large
as $ 4 \,GeV^2$ \cite{KL}.

 We hope that the simultaneous analysis of all HERA
data , which we plan
 to undertake, will   be consistent only with the  saturation approach. In
doing so we
have to rely on the hdQCD evolution equations rather then on models.

\section{Predictions for THERA:}
 As we have pointed out it is  rather dangerous  to discuss
predictions for lower $x$ in our model, since the model should be replaced
by the solution of the correct evolution equation of \eq{GLRINT}. However, 
in this section we summarize some of our predictions for  as  low $x$ as
$ x
\rightarrow 10^{-6}$ since this region of $x$ is under detailed
discussion  due to plans of the THERA extension of TESLA project
\cite{THERA}. We would like to recall that our predictions underestimate 
the solution to the nonlinear evolution equation, but we can use them as a
first estimate in our  search for  collective phenomena in high
parton
density system. 

~

\centerline{\bf The unitarity boundary and THERA kinematic region:}

~

We start from the prediction which, in principle, does not depend on the
exact form of the correct evolution equation, namely, from the unitarity
boundary for the $F_2$ and $xG(x,Q^2)$ structure functions. This problem
has been considered in Ref. \cite{AGLFRST} and it turns out that we have
the following unitarity boundaries for $F_2$ and $xG(x,Q^2)$:
\be
\frac{\partial F_2(x,Q^2)}{\partial \ln Q^2}\,\,\,&<&\,\,\,\frac{1}{3
\,\pi^2}\,\,Q^2\,R^2\,\,;\label{UB1}\\
\frac{\partial^2 xG(x,Q^2)}{\partial \ln Q^2\,\,\partial \ln(1/x)}
\,\,\,&<&\,\,\,\frac{2}{\pi^2}\,\,Q^2\,R^2\,\,;\label{UB2}
\ee
where $R$ is the nucleon radius. Actually, in \eq{UB1} and \eq{UB2} we
need to $R^2 = < b^2_t>_{F_2}$ and $R^2 = < b^2_t>_{xG}$, respectively
(see Fig. 7). As it shown in Fig.7, the interaction radii depend on energy
and $Q^2$ but the  dependence is so weak that we can neglect it in the
first approximation. 

\begin{figure}
\begin{center}
\psfig{file=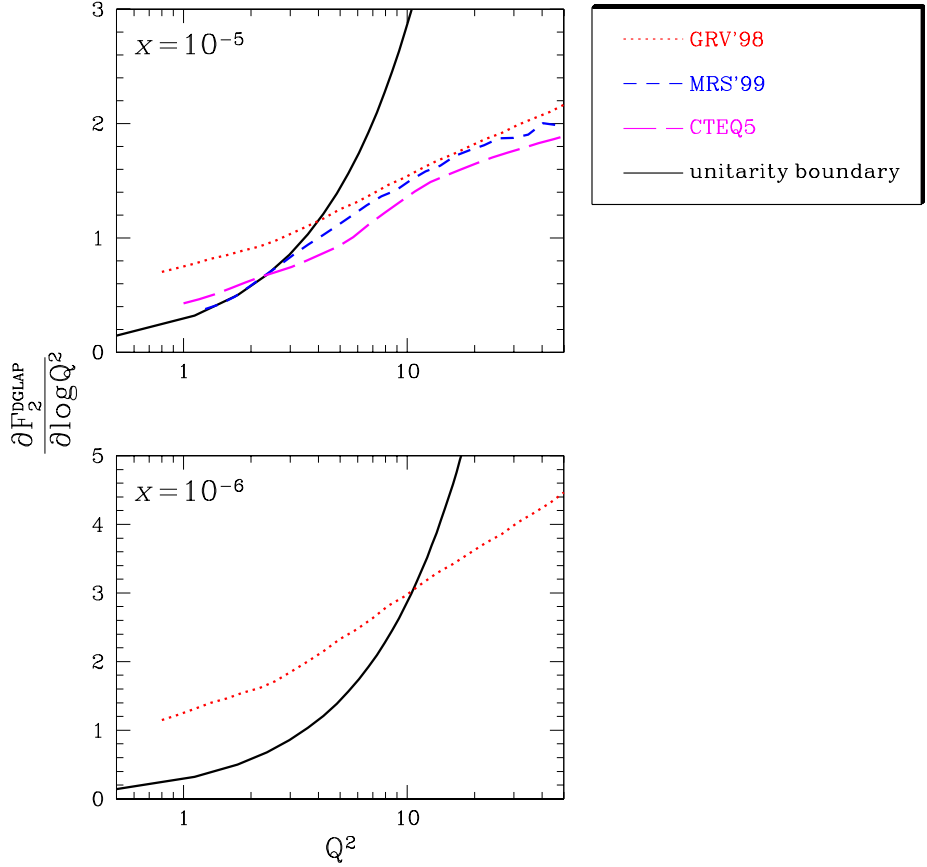,width=13cm}
\end{center}
\caption{\it The unitarity boundary for $F_2$-slope and predictions in
GRV,MRS and CTEQ parameterizations for the solution of the DGLAP
evolution equations  in THERA kinematic region. }
 \label{unbf2}
 \end{figure}

In Fig.~\ref{unbf2} we plot \eq{UB2} and the predictions for
$F_2$-slopes in GRV\cite{GRV},in  MRS \cite{MRS} and in CTEQ \cite{CTEQ}
parameterizations. One can see that the $F_2$-slope in the DGLAP evolution
equations reaches the unitarity boundary at $Q^2  \leq 3\,GeV^2 $  for
$x=10^{-5}$ and even for $Q^2  \leq 10\,GeV^2 $ for $x=10^{-6}$.

\begin{figure}
\begin{center}
\psfig{file=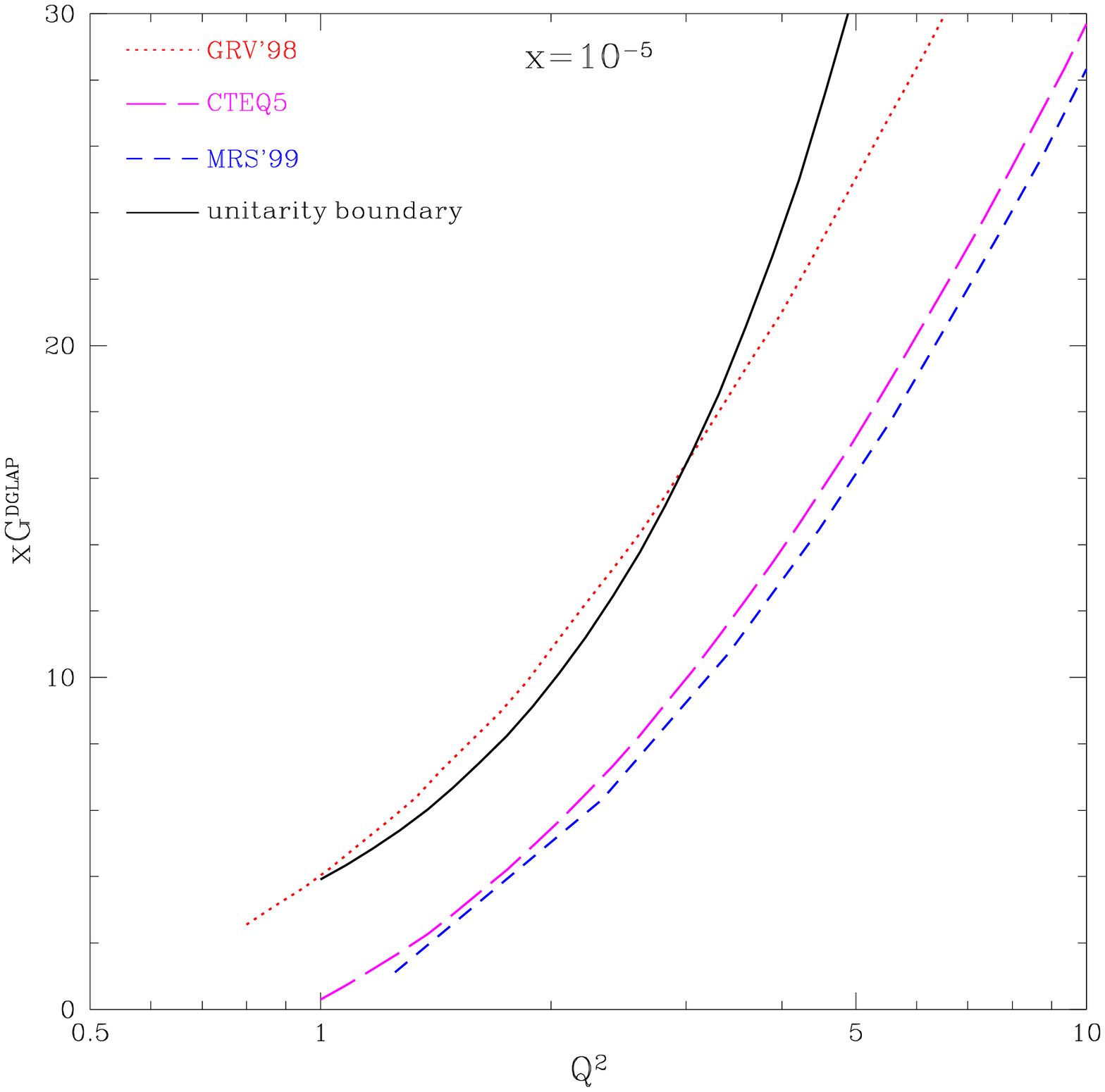,width=13cm}
\end{center}
\caption{\it The unitarity boundary for $xG(x,Q^2)$  and predictions in
GRV,MRS and CTEQ parameterizations of the solution of the  DGLAP
evolutions equations in THERA kinematic region. }
 \label{unbg}
 \end{figure}

Fig.~\ref{unbg} shows the comparison of the DGLAP evolution equations
with the unitarity boundary for the case of gluon structure function. In
Fig.~\ref{unbg}
\eq{UB2} is  modified using the DGLAP evolution equation in the region of
low $x$, namely,
\beq \label{DLAG}
\frac{\partial^2 xG^{DGLAP}(x,Q^2)}{\partial \ln Q^2\,\,\partial \ln(1/x)}
\,\,\,=\,\,\frac{N_c \alpha_S(Q^2)}{\pi}\,\,\, xG^{DGLAP}(x,Q^2)\,\,.
\eeq
Using \eq{DLAG} one can rewrite \eq{UB2} in the form
\beq \label{UB3}
xG^{DGLAP}(x,Q^2)\,\,\,<\,\,\,\frac{2}{\pi\,N_c\,\alpha_S(Q^2)}\,\,Q^2\,R^2\,\,.
\eeq

Based on  Fig.~\ref{unbg} we are not  optimistic of finding gluon
saturation for the unitarity constrains on the gluon structure function
in THERA kinematic region. The $F_2$-slope at lower $x$ ($x \approx
10^{-6}$)  will give more reliable limit on the gluon saturation.

\begin{figure}
\begin{center}
\epsfxsize=10cm
\leavevmode
\hbox{ \epsffile{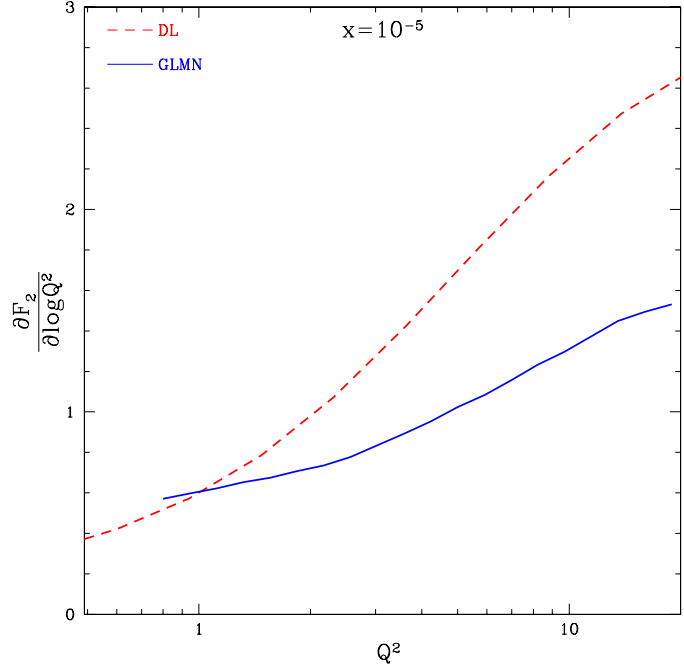}}
\end{center}
\caption{\it The prediction for the $F_2$-slope in our approach (GLMN) and
in Donnachie-Landshoff approach (DL).}
\label{slpt}
\end{figure}

However, the  calculations, given in Fig. 6 using both our model and the
solution to
the \eq{GLRINT},  suggest that the corrections to at least $xG$ is 
rather large in the THERA kinematic region.

$$\mathbf{ \frac{\partial F_2}{\partial \ln Q^2}}$$
The THERA kinematic region provides a new opportunity to check the value
of
SC using the $F_2$ slope. At least, the THERA data would help us to
differentiate between two approaches: our approach, based on the gluon
saturation at low $x$ and the approach based on the matching between
``soft" and ``hard" interaction with a sufficiently large momentum scale
for ``soft"  contribution.
Fig. ~\ref{slpt} shows the predictions for the $F_2$-slope in our approach
(  GLMN in figure ) and the Donnachie-Landshoff approach\cite{DL2P}. The
difference
is rather large and can be measured.

~

\centerline{\bf J/$\mathbf{\Psi}$ production:}

~

\begin{figure}
\begin{tabular}{c c}
\psfig{file=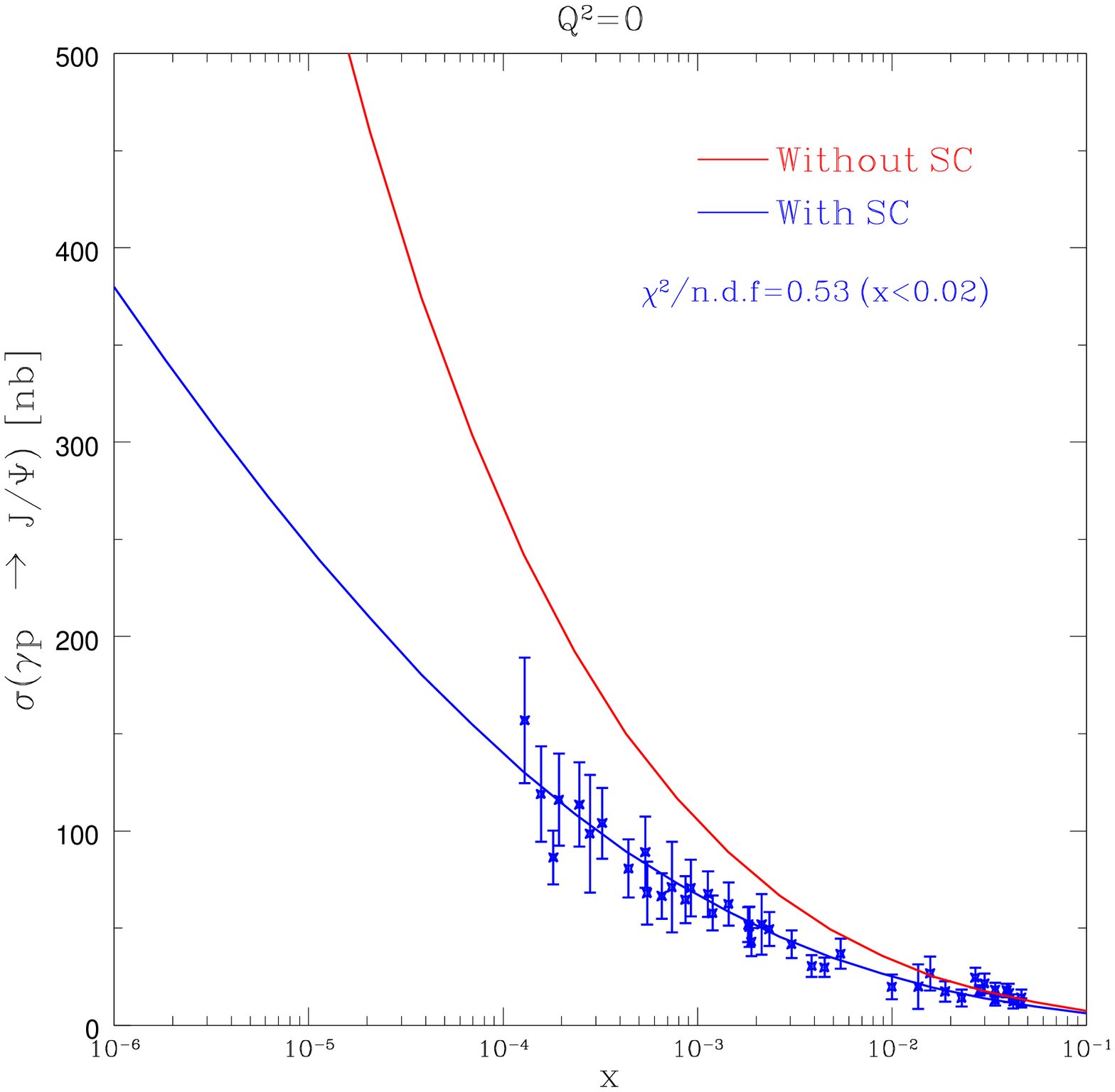,width=8cm,height=8cm}&
\psfig{file=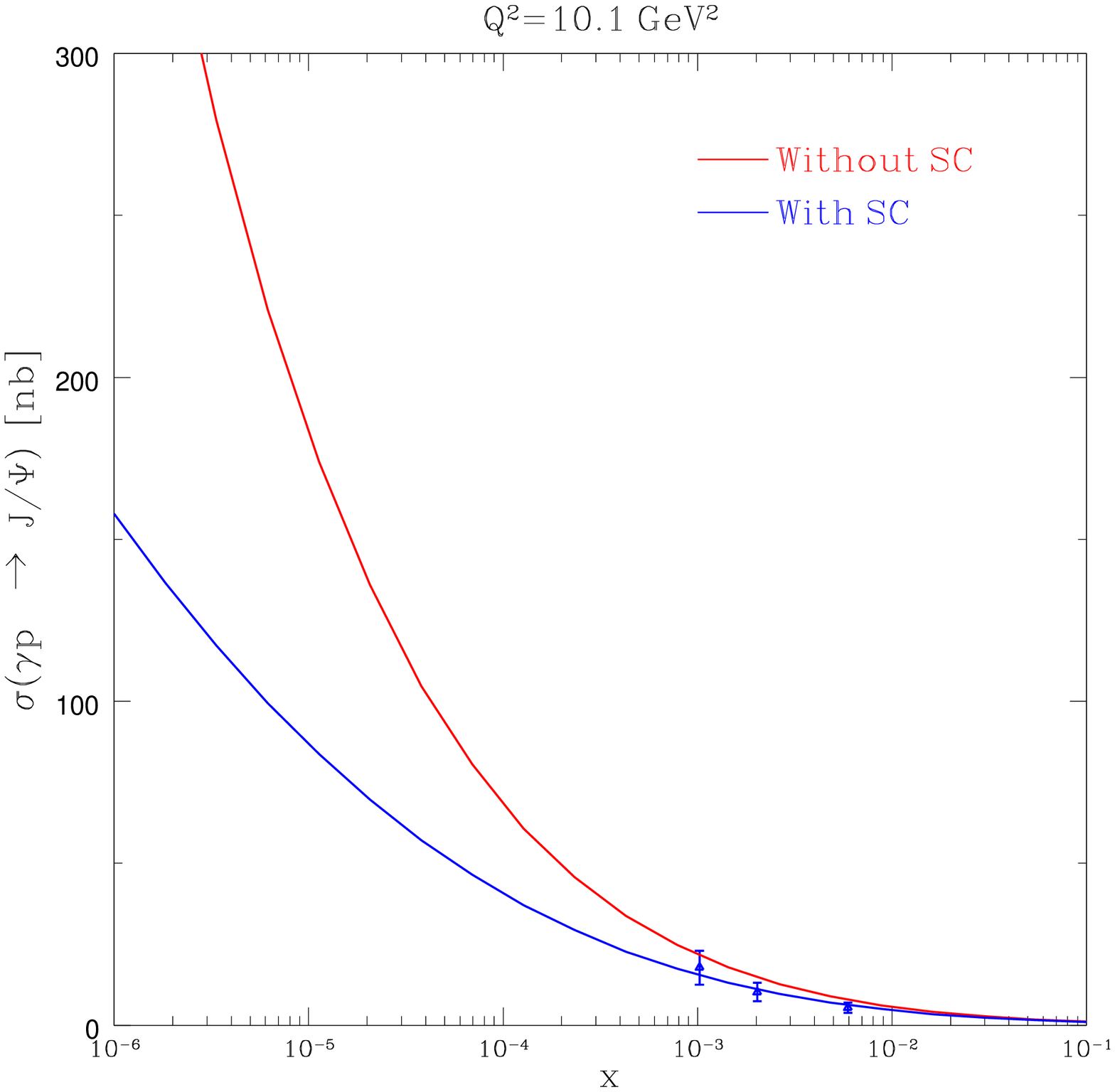,width=8cm,height=8cm}\\
\end{tabular}
\caption{\it Predictions for J/$\Psi$ production in THERA kinematic
region together with available experimental data.}
 \label{psiprod}  
 \end{figure}

Fig.~ \ref{psiprod} shows that the difference between our predictions and
the DGLAP increase at the THERA kinematic region, which leads us to
speculate  that we will see  gluon saturation at THERA.

~

\centerline{\bf Maxima in the ratios and the higher twist contributions}

~

Fig.~\ref{max} shows that we expect a characteristic behaviour of the
ratio
$F^D_L/F^D_T$ as a function of $Q^2$. It turns out to be wider than at
higher $x$ and the maximum occurs at the value of $Q^2_{max} \approx 6
\div 7 \,GeV^2$. Such a large value of $Q^2_{max}$ makes the calculations
reliable and we expect that the measurement of this observable in the  
THERA
kinematic region will help us to extract the value of the saturation
scale from the experimental data.  In THERA kinematic region ( at
$x\approx 10^{-5}$ ) we expect that the higher twist contributions
to be of the same order as the leading twist ones at sufficiently high
value of $Q^2$, namely, for $ F_L$ and for $F^D_T$ this value of $Q^2$ is
about $ 5 \div 7 \,GeV^2$ while for $F^D_L$ it is even larger, about $ 15
\,GeV^2$.  The fact that the higher twist contributions become visible at
large value of $Q^2$ lead to a visible violation of the usual DGLAP
evolution approach which should  be possible to measure.

\section*{Acknowledgments}

The authors are very much indebted to our coauthors and friends with
whom we discussed our approach on  a everyday basis Ian Balitsky,Jochen
Bartels ,
Krystoff Golec Biernat,Larry
McLerran, Dima Kharzeev, Yuri Kovchegov and  Al Mueller for their help and
fruitful discussions on the subject. E.G. ,  E. L.  and U.M. thank BNL
Nuclear
Theory Group and  DESY Theory group
for their hospitality and
creative atmosphere during several stages of this work.

This paper is the fruit  of the useful discussions during
Amirim meeting and it is a pleasure for us to thank all participants of
the Amirim workshop:  H.~Abramowicz,
J.~Bartels,
L.~Frankfurt, K.~Golec-Biernat, 
E.~Gurvich, H.~Jung,  S.~Kananov, H.~Kowalsky, A.~Kreisel, 
A.~Levy, P.~Marage, M.~McDermott, 
 P.~Saull, S.~Schlenstedt,  G.~Shaw,  M.~Strikman, J.~Whitmore .

This research was supported in part by the BSF grant $\#$
9800276 and by
Israeli Science Foundation, founded by the Israeli Academy of Science
and Humanities.

 \end{document}